\documentclass[useAMS,usenatbib]{mnras}
\def\draftversion{0} 
\usepackage{amssymb,latexsym,graphicx,natbib,eufrak,times,amsmath,xspace,ifthen}

\ifthenelse{ \draftversion > 0 }{
} {
  
}

\ifthenelse{\equal{\draftversion}{1}}{
  \usepackage{xcolor}
  \newcommand{\sep}[1]{\par\begin{color}[rgb]{0,0.4,0}\begin{center}\hrule\end{center}\end{color}\par} 
  \newcommand{\todo}[1]{\begin{color}{red}\ \ifthenelse{\equal{#1}{}} {$\bullet\bullet\bullet$} {$\bullet$\ #1 $\bullet$}\end{color}} 
  \newcommand{\idea}[1]{\begin{color}[rgb]{0,0.4,0}\textit{#1}\end{color}} 
  \newcommand{\sk}[1]{\begin{color}[rgb]{0.6,0,0.6}#1\end{color}} 
  \newcommand{\toc}{\par\begin{color}[rgb]{0.6,0,0.6}\begin{center}\hrule\vspace{0.5mm}\begingroup\small\let\cleardoublepage\relax\let\clearpage\relax\mytoc\endgroup\vspace{0.5mm}\hrule\end{center}\end{color}\par} 

  }{
  \newsavebox{\trashcan}

  \newcommand{\sep}[1]{}
  \newcommand{\todo}[1]{}
  \newcommand{\idea}[1]{}
  \newcommand{\sk}[1]{}
  \newcommand{\toc}{}

  }
 \makeatletter\newcommand\mytoc{\@starttoc{toc}}\makeatother 
\long\def\symbolfootnote[#1]#2{\begingroup%
\def\thefootnote{\fnsymbol{footnote}}\footnote[#1]{#2}\endgroup} 

\newcommand{\fig}[2][]{Figure#1~\ref{fig:#2}}

\newcommand{\sect}[2][]{Section#1~\ref{sec:#2}}
\newcommand{\app}[2][]{Appendix#1~\ref{sec:#2}}

\newcommand{\bb}[1]{\ifmmode \mbox{\boldmath $ #1$} \else  \mbox{\boldmath $#1$} \fi}

\newcommand{\mh}{\ensuremath{\textrm{\,--\,}}}    
\newcommand{\U}[1]{\ensuremath{\mathrm{~#1}}}     
\newcommand{\e}[1]{\ensuremath{\times 10^{#1}}}   

\newcommand{\yr}{\U{yr}}
\newcommand{\Myr}{\U{Myr}}          
\newcommand{\Gyr}{\U{Gyr}}          \newcommand{\gyr}{\Gyr}
\newcommand{\pc}{\U{pc}}
\newcommand{\kpc}{\U{kpc}}
          
\newcommand{\Msun}{\U{M}_{\odot}}   \newcommand{\msun}{\Msun}
\newcommand{\Msunyr}{\Msun\yr^{-1}} 
\newcommand{\Zsun}{\U{Z}_{\odot}}   
\newcommand{\cc}{\U{cm^{-3}}}
\newcommand{\K}{\U{K}}
\newcommand{\Msunpc}{\U{M_{\odot}\ pc^{-3}}}    \newcommand{\msunpc}{\Msunpc}
\newcommand{\kms}{\U{km\ s^{-1}}}

\newcommand{\dex}{\U{dex}}

\newcommand{\tdep}{\ensuremath{t_\mathrm{dep}}\xspace}        
\newcommand{\afe}{\ensuremath{[\alpha/\mathrm{Fe}]}\xspace}       
\newcommand{\feh}{\ensuremath{[\mathrm{Fe/H}]}\xspace}       

\newcommand{\ramses}{{\small RAMSES}\xspace}

\newcommand{\hop}{{\small HOP}\xspace}

\newcommand{\vintergatan}{{\small VINTERGATAN}\xspace}

\newcommand{\lund}{Department of Astronomy and Theoretical Physics, Lund Observatory, Box 43, SE-221 00 Lund, Sweden}
\newcommand{\surrey}{Department of Physics, University of Surrey, Guildford, GU2 7XH, UK}

\defcitealias{Agertz2020}{Paper I}
\defcitealias{Renaud2020b}{Paper III}

\title[VINTERGATAN II]{VINTERGATAN II: the history of the Milky Way told by its mergers}
\author[Renaud et al.] {Florent~Renaud$^1$\thanks{florent@astro.lu.se}, Oscar~Agertz$^{1}$, Justin~I.~Read$^{2}$, Nils~Ryde$^1$, Eric~P.~Andersson$^{1}$, \newauthor
Thomas~Bensby$^{1}$, Martin~P.~ Rey$^{1}$ and Diane~K.~Feuillet$^{1}$\\
$^1$ \lund\\
$^2$ \surrey}

\date{Accepted 2021 January 25. Received 2021 January 18; in original form 2020 June 12}

\begin{document}
\maketitle


\begin{abstract}
Using the \vintergatan cosmological zoom simulation, we explore the contributions of the in situ and accreted material, and the effect of galaxy interactions and mergers in the assembly of a Milky Way-like galaxy. We find that the initial growth phase of galaxy evolution, dominated by repeated major mergers, provides the necessary physical conditions for the assembly of a thick, kinematically hot disk populated by high-\afe stars, formed both in situ and in accreted satellite galaxies. We find that the diversity of evolutionary tracks followed by the simulated galaxy and its progenitors leads to very little overlap of the in situ and accreted populations for any given chemical composition. At a given age, the spread in \afe abundance ratio results from the diversity of physical conditions in \vintergatan and its satellites, with an enhancement in \afe found in stars formed during starburst episodes. Later, the cessation of the merger activity promotes the in situ formation of stars in the low-\afe regime, in a radially extended, thin and overall kinematically colder disk, thus establishing chemically bimodal thin and thick disks, in line with observations. We draw links between notable features in the \feh - \afe plane with their physical causes, and propose a comprehensive formation scenario explaining self-consistently, in the cosmological context, the main observed properties of the Milky Way.
\end{abstract}

\section{Introduction}

Stellar populations encode a wealth of information about the assembly history of their host galaxies. For instance, the evolutions of the conditions of star formation can be explored through the chemical composition of the stars. In the Milky Way, stars gather in two sequences, at high and low \afe \citep{Bensby2005, Nissen2010, Anders2014, Hayes2018}. The high-\afe sequence corresponds (at least approximately) to old stars (e.g. \citealt{Haywood2013, Bensby2014, Feuillet2019, Ciuca2020}) that are kinematically hot (e.g. \citealt{Nordstrom2004, Hayden2020}) and geometrically distributed in a thick \citep{Yoshii1982, Gilmore1983} and radially compact disk \citep{Bensby2011, Minchev2015, Bovy2016}. Conversely, the low \afe component shows the opposite characteristics: younger stars with low \afe in a kinematically cold, thin and radially extended disk (see \citealt{Kawata2016} and \citealt{Bland2016} for reviews). However, the physical origins of these populations are still unclear, and it is not yet established whether the distinctions in age, kinematics, chemistry, and geometry share the same physical origins or not.

To explain the formation of two chemically distinct sequences, \citet{Chiappini1997} proposed the two-infall scenario, in which the Milky Way first self-enriches for $1 \Gyr$, before the accretion of low-metallicity gas resets the chemical evolution of the galaxy by lowering \feh. In a revised version \citep{Spitoni2019}, the second infall takes place later ($4.3 \gyr$), such that the Milky Way has time to populate the high-\afe sequence first, before the chemical reset allows for the formation of the low-\afe branch. 
To better reproduce the deficit of stars with intermediate-\afe compositions, and thus to more strongly mark the chemical bimodality, the chemical evolution models of \citet{Lian2020} invoke a more abrupt transition from the high to low-\afe regime, with a early and rapid quenching of the first formation episode. In such scenarios, the assembly of the stellar populations is thus a sequential process, i.e. with no simultaneous events. Similar scenarios, in particular invoking a discontinuity in the gas accretion history \citep{Chiappini2001, Birnboim2007, Bekki2011}, have been found in cosmological zoom-in suites of simulations AURIGA \citep{Grand2018} and NIHAO \citep{Buck2020b} which report the presence of a chemical bimodality in a significant fraction of the cases explored.

However, other simulations have reached different conclusions on the existence or likelihood of the \afe bimodality. For instance, with a cosmological zoom simulation in the FIRE framework, \citet{Ma2017} reported the absence of bimodality. At lower resolution but over a larger sample, \citet{Mackereth2018} found that the bimodal nature of the \afe distribution in the Milky Way was only very rarely reproduced in the EAGLE simulation, and concluded that the Galaxy must have experienced a very peculiar assembly history. The diversity of conclusions from cosmological simulations reveals a large sensitivity to the implementations of physical processes. The adequation of the recipes of star formation and feedback with the adopted resolution seems particularly crucial in this field \citep{Buck2019}.

With the setup of a disk galaxy in isolation, \citet{Clarke2019} showed that the chemical bimodality can result from evolving physical conditions of star formation along galaxy formation. High gas fractions allow for large-scale instabilities and the formation of massive gas clumps in which rapid star formation leads to enhanced \afe (see also \citealt{Noguchi1998, Bournaud2009b}). Dynamical scatter by the massive clumps would then explain the correspondence between high \afe, kinematically hot, and geometrically thick properties. However, in the simulation of \citet{Clarke2019}, these clumps arise from a pre-existing thin disk, which contradicts the observations of an older population at high rather than low \afe \citep{Feuillet2019, Ciuca2020}. With a comparable setup of isolated disks, \citet{Khoperskov2020} concluded that the bimodality results from an intrinsic slowing down of the star formation activity, from a compact turbulent disk to a more quiescent thin disk. Other scenarios also mention the origins of the thick-thin dichotomy as a purely internal effect, for instance induced by radial migration \citep{Schoenrich2009, Loebman2011, Frankel2018}, although this mechanism alone is not sufficient to thicken the disk \citep{Minchev2012}.

In the family of external factors, the role of galaxy interactions and mergers remains highly uncertain and is poorly constrained observationally. While the accretion of low-mass satellites at low redshift is traced by the stellar streams they leave (e.g. \citealt{Ibata2001, Belokurov2006}), the impact of major mergers is, counter-intuitively, less readily traceable. The reason for this is the stronger dynamical friction between extended galaxies that accelerates their coalescence and mixing with in situ stars in more central volumes. \citet{Helmi2018} and \citet{Belokurov2018} have identified two possible signatures of a major accretion event in the Gaia data (potentially the last one the Milky Way experienced) now called the Gaia-Enceladus-Sausage (see \citealt{Helmi2020}). However, the details of the interaction, of the progenitor galaxy, and how the Milky Way responded this event remain to be established.

Galaxy mergers in general have been suspected to play a central role in building the thick disk of the Galaxy. For instance, the majority of, perhaps even all, the thick disk stars could result from the accretion of gas-rich galaxies \citep{Brook2004}, and/or the tidal disruption of incoming satellites and the accretion of their stars \citep{Abadi2003, Read2008}. These same mergers could tidally heat and thicken a pre-existing disk \citep{Quinn1993, Kazantzidis2008, Read2008, Villalobos2010}, and trigger important episodes of star formation \citep{Gallart2019, Ruiz2020}. Mergers have also been proposed to explain the formation of the stellar halo, either through a limited number of major events (including the Gaia-Enceladus-Sausage, \citealt{Myeong2019}), or with a Gaia-Enceladus-Sausage-like event coupled with a larger number of accreted smaller progenitors \citep{Monachesi2019}. However, \citet{Ruchti2010} found that the chemical composition of halo stars is incompatible with a formation of the halo dominated by the accretion of dwarf satellites similar to those observed in the Local Group at the present-day. Finally, from Gaia data, \citet{Haywood2018} suggested that halo-like dynamics could originate from a combination of accretion and heating of the thick disk by mergers (see also \citealt{DiMatteo2019}).

Using the \vintergatan simulation introduced in \citet{Agertz2020} (hereafter \citetalias{Agertz2020}), we present here an analysis of the roles of secular evolution, galaxy interactions and mergers in the assembly of a Milky Way-like galaxy\footnote{Movies are available at:\\ \url{http://www.astro.lu.se/~florent/vintergatan.php}}. We focus on the evolution of the chemical composition of the stellar populations, and the transition between the thick and the thin disks. The numerical method is summarized from \citetalias{Agertz2020} in \sect{method}. \sect{results} gathers multi-dimensional diagnostics applied to the simulation to infer the role of mergers. The results are discussed in \sect{discussion}, and \sect{scenario} proposes a comprehensive formation scenario of Milky Way-like galaxies, replaced in the broader context of galaxy formation and evolution.

\section{Method}
\label{sec:method}

We use the \vintergatan cosmological zoom simulation of the formation of a Milky Way-like galaxy presented in \citetalias{Agertz2020}, and briefly summarize the numerical method below. We use the adaptive mesh refinement code \ramses \citep{Teyssier2002} and the cosmological zoom-in technique \citep{Hahn2011} on the initial conditions ``m12i'' from \citet{Hopkins2014}, i.e. a dark matter halo of $M_{200} = 1.3 \e{12} \Msun$ and $R_{200} = 334 \kpc$ at $z=0$, the same as in \citet{Renaud2017}. The resolutions in the zoom volume are $\approx 3.5 \e{4} \Msun$ for the dark matter particles and $\approx 7 \e{3} \Msun$ for the gas, and a spatial resolution of $\approx 20 \pc$ in the dense interstellar medium (ISM). Due to the high numerical cost, the simulation is stopped at $z=0.19$ ($\approx 2.3 \Gyr$ ago). At this point, the \vintergatan galaxy reaches a stellar mass of $6\e{10} \msun$ within $20 \kpc$. Since the main structures are in place and the galaxy evolves in a secular manner, we do not expect any violent effect to occur in the remaining time, and use our last snapshot as a proxy for present-day conditions.

The initial gas metallicity is set to $10^{-3}\Zsun$ to mimic the enrichment from the unresolved population III stars \citep{Wise2012}. Gas cooling follows metallicity-dependent tabulated functions from \citet{Sutherland1993} and \citet{Rosen1995} for atomic and molecular cooling respectively. An ultraviolet background produces heating (assuming a re-ionization redshift of 8.5, \citealt{Haardt1996}). Star formation proceeds at a local star formation efficiency (as in \citealt{Padoan2012}) per free-fall time above a density threshold of $100 \cc$, and below a temperature of $100 \K$. Star particles of $10^{4}\Msun$ then represent single-age populations. The stellar feedback model includes stellar winds, radiative pressure and the local resolution-dependent injection of energy and momentum from type-II and type-Ia supernovae (SNe), following \citet{Agertz2013}.

The SNe release oxygen and iron that are tracked using passively advected scalars. We compute the abundance ratios \afe and \feh using the solar mass fraction of 0.0097 for oxygen and 0.00185 for iron from \citet{Anders1989}. As in \citetalias{Agertz2020}, and because of uncertainties on chemical enrichment models, we consider the $\alpha$-elements to be fully traced by the oxygen abundance. Considering the contribution of other elements (e.g. through a weighted average as in \citealt{Bovy2016}) and/or different models for yields would add a normalization factor which would shift the values of \afe and \feh in our results. Instead of seeking which model would give the best match with observations, we choose to use and present the raw data from the simulation and limit our analysis to relative comparisons of chemical features rather than providing absolute quantities. 

The galaxies are identified with a clump-finder algorithm run on the stellar component (using the code \hop, \citealt{Eisenstein1998}, with the peak, saddle and outer densities of $10^{-3}$, $5\e{-4}$, $10^{-7} \msunpc$, respectively) at each snapshot, i.e. with a time resolution of $\sim 150 \Myr$. In the following, we refer to ``\vintergatan'' as the most massive progenitor of the most massive galaxy in the final snapshot, i.e. the Milky Way-like galaxy. The plane of the disk of \vintergatan is defined as the plane orthogonal to the total angular momentum vector of all its stars, and passing through the stellar center of mass.

In the rest of the paper, the term ``in situ'' refers to the stellar material formed in the most massive galaxy progenitor, while ``accreted'' refers to stellar material that forms in other galaxies, which are then accreted onto \vintergatan. To ease the comparison with stellar ages, all times are given as lookback times (i.e. with $t=0$ corresponding to present-day). Finally, in the following, the term ``metallicity'' refers to the \feh abundance ratio.

\section{Results}
\label{sec:results}

\subsection{Star formation history}

\vintergatan is detected as early as star formation starts ($z \approx 18$), and experiences 6 major mergers (arbitrarily defined as stellar mass ratios greater than 1:10) at redshifts 5.7, 4.2, 2.7 (2 events), 2.2 and 1.2, corresponding to lookback times of 12.8, 12.3, 11.3 (2 events), 10.7 and $ 8.7\Gyr$.

\begin{figure}
\centering
\includegraphics{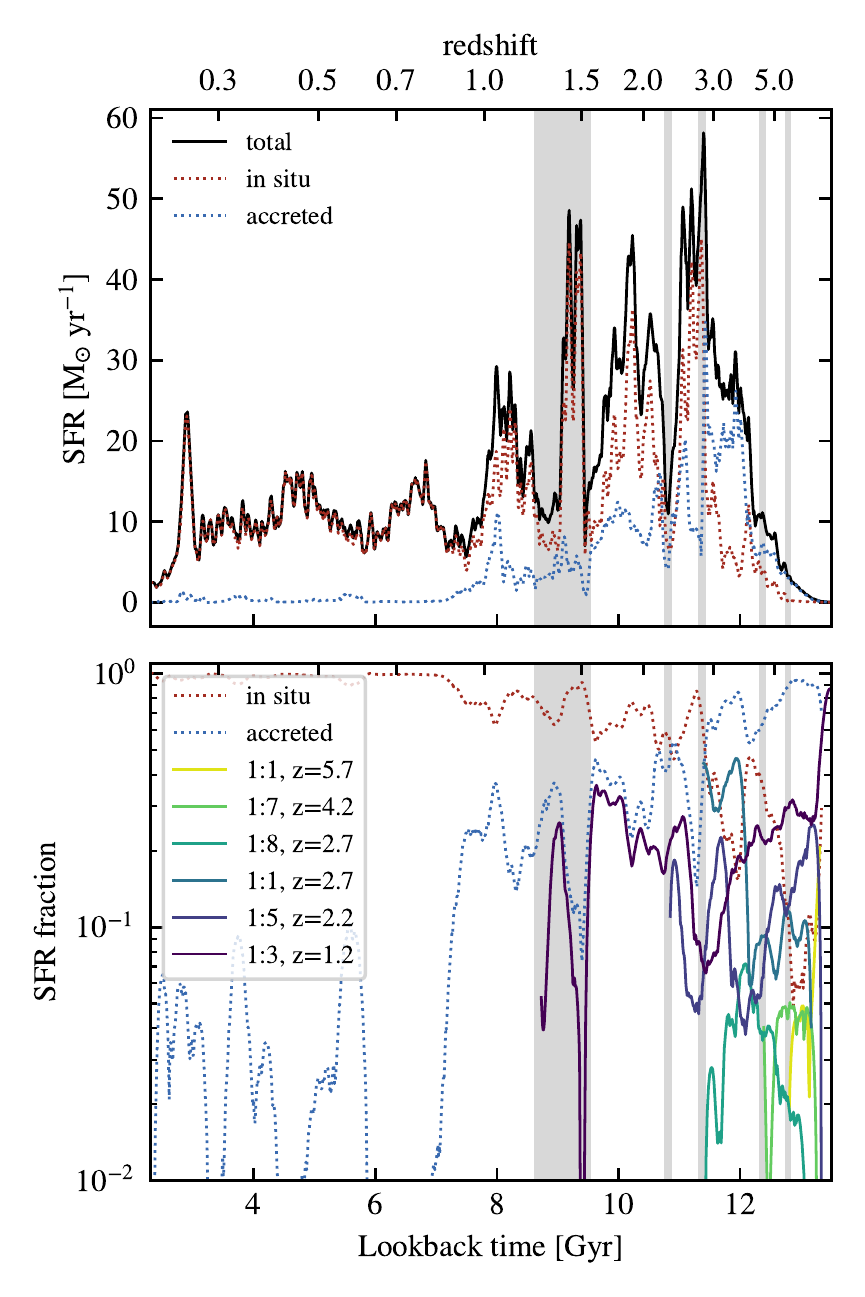}
\caption{Top: evolution of the star formation rate in \vintergatan, and the contribution of in situ and accreted stars. Bottom: relative contributions to the total SFR of the in situ and accreted populations, and of the individual galaxies involved in major merger events. The stellar mass ratios and the redshifts of the mergers are indicated in the legend. The curves have been smoothed with a Savitzky-Golay algorithm for the sake of clarity. Shaded areas indicate the epochs of major mergers (from first pericenter passage to final coalescence). Apart from localized peaks linked with starburst activity during merger events, the contribution of the accreted population steadily declines until the final epoch probed, and becomes sub-dominant at $z = 2.8$ ($11.4 \Gyr$ ago).}
\label{fig:sfr}
\end{figure}

\fig{sfr} shows the evolution of the star formation rate (SFR) of \vintergatan, measured as the distribution of the ages of the stars in the galaxy in the last snapshot, weighted by their mass at birth\footnote{The peak at $\sim 3\Gyr$ is caused by a numerical artifact, as the maximum resolution of the adaptive refinement increases, leading to a weaker pressure support in the densest gas. The amount of stars formed during this event is negligible with respect to the entire galaxy, and this does not affect our conclusions.}. The relative contribution of the in situ and accreted population reveals, as expected, that the in situ formation largely dominates at late times, while the accreted populations constitute the main contributors to stars older than $\approx 11.4 \Gyr$ ($z \gtrsim 2.8$).

The individual contributions to the SFR of the galaxies involved in major mergers are shown in \fig{sfr} (bottom panel). The most massive companions host an enhancement of their SFR a few 100 Myr before they merge, which indicates a starburst activity induced at an early stage of interaction (e.g. at the first passage, see \citealt{Renaud2014b}). However, the low mass galaxies, in which gas reservoirs are more sensitive to tidal and ram pressure stripping, suffer instead from a sharp decline of their SFRs (known as fast quenching, e.g. \citealt{Farouki1981, Moore1996}). The accreted galaxies individually contribute to up to $30 \mh 40\%$ of the total SFR until $z \approx 1$ ($8 \Gyr$ ago), and about $10\mh 30\%$ for the galaxy involved in the last major merger alone. At later times, the SFR of the accreted population (hence exclusively from late minor mergers) drops to a level of a few percent, with sparse peaks reaching $\sim 10\%$, and each lasting about $100 \Myr$. 

\fig{sfr} shows that, at all epochs, high levels of contribution of the accreted population to the SFR originate from galaxies that formed at the earliest epochs ($z \gtrsim 5$). This is because such galaxies have more time than others to accrete material themselves (through cold streams and/or mergers), thus grow in mass and become less fragile to tidal stripping and quenching \citep{Simpson2018} when they eventually merge with \vintergatan. This also implies that they have the opportunity to form and retain more chemically enriched material than the lower mass, younger galaxies, which impacts the chemical properties of their stars, as discussed below.

The slow but long-lasting growth of \vintergatan after the major merger-dominated phase ($\lesssim 9 \Gyr$, $z \lesssim 1.2$) reduces the relative contribution of these galaxies. The entire accreted population constitutes $28\%$ of the final stellar mass of \vintergatan. The individual contributions of the major mergers account for 0.02, 0.1, 0.6, 5.2, 2.8, and 10.5\% of the final stellar mass of the galaxy respectively (in chronological order). In other words, the 6 major mergers combined contribute $20\%$ of the final stellar mass, half of this from the last event only.

\subsection{Chemistry of inflows and outflows}
\label{sec:inflows}

\begin{figure}
\centering
\includegraphics{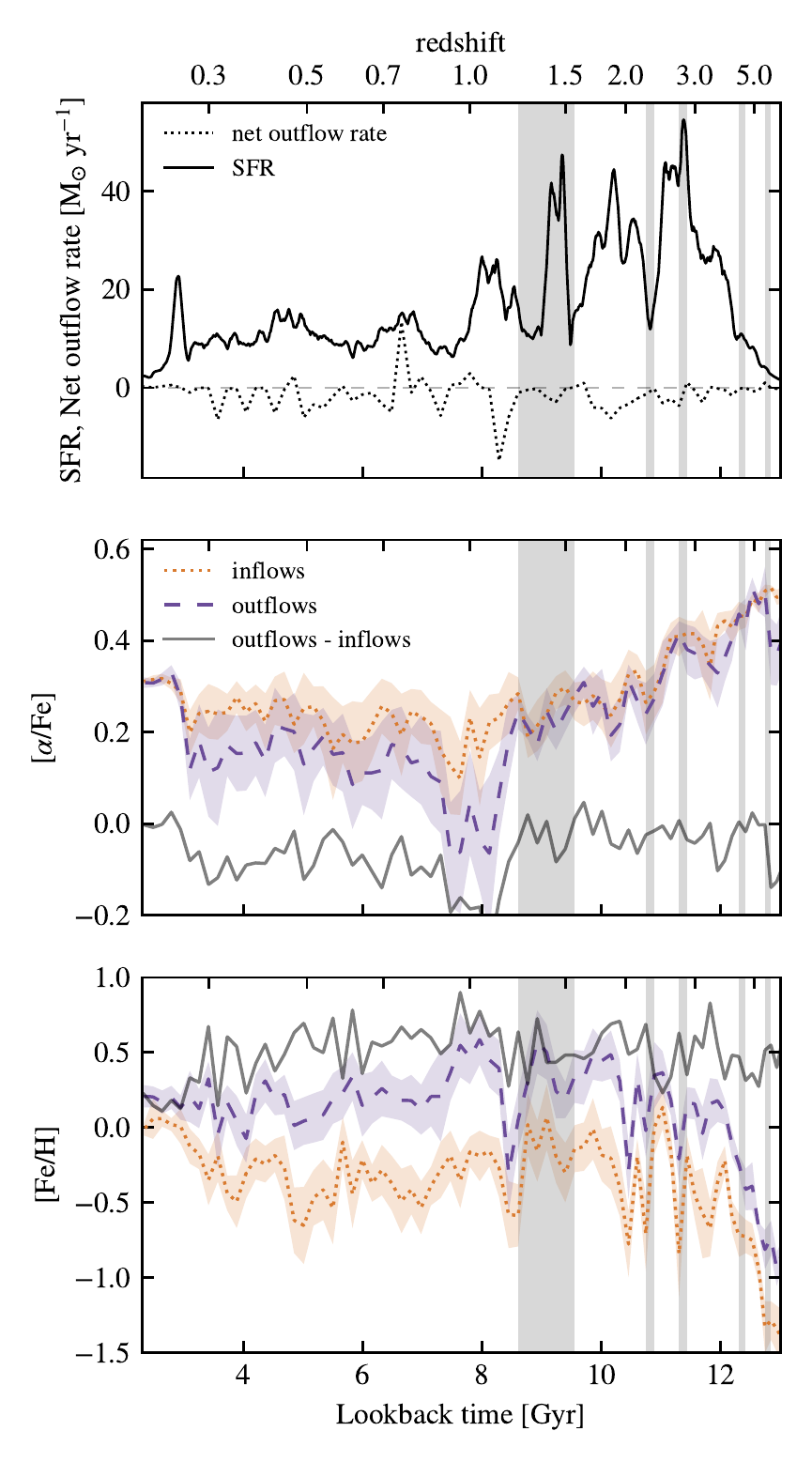}
\caption{Top: net outflow rate, as the difference between the outflowing and the inflowing gas masses. The SFR is repeated from \fig{sfr}, for reference. Gray shaded areas show the epochs of major mergers. Middle and bottom: evolution of the chemical compositions of inflowing and outflowing gas (see \sect{inflows} for details). The colored shaded areas indicate the standard deviation of the abundances, across all directions from/to the galaxy, and the gray solid lines show the difference between the median outflows and inflows. The chemical evolution of in/outflowing gas is complex and rapidly fluctuating, but major mergers seem to have the strongest influence.}
\label{fig:outflows}
\end{figure}

In addition to the hierarchical assembly through mergers, the other key component in the building of a galaxy is the accretion of gas via cold flows, and its ability to eject (or retain) its enriched material via feedback. \fig{outflows} shows the evolution of the chemical composition of the inflowing and outflowing gas. In/outflows are computed using the radial velocity of the total gas (as the projection of the velocity vector onto the radial vector from the center of mass of the galaxy) in concentric spherical shells of inner radius 2, 3 and 5 times the (evolving) stellar half-mass radius of \vintergatan and with a thickness corresponding to 10\% of this radius. The data shown in \fig{outflows} is the median of the values measured in these three shells, in each snapshot. The outflows comprise galactic winds launched by feedback and the ejection of tidal debris during interactions. Conversely, the inflows consist of a fraction of these ejecta falling back onto the galaxy, and the accretion of companion galaxies and intergalactic gas. Star formation proceeds in the gas from the inflows, but also from the recycling of stellar feedback: about $40\mh 50\%$ of the stellar mass is retroceded to the ISM over Gyr-long timescales. This explains why the SFR can be higher than the net inflow rate \citep{Leitner2011}. The net budget thus results from a complex interplay of many processes, several being active simultaneously at the same location, or in different directions around the galaxy. A detailed analysis of the contributions of individual processes is very involved and beyond the scope of this paper.

The net outflow rate (i.e. outflows -  inflows) is negative on average, which accounts for the overall mass growth of the galaxy. Mergers correspond to a rapid inflow (i.e. a sudden drop of the net outflow rate), partly balanced by the outflows launched by feedback from the associated starburst episodes. A delay between such inflows and outflows is due to molecular cloud assembly, their compression and collapse, star formation, the onset of supernova feedback and its propagation to large radii (as particularly visible immediately after the last major merger at $z \approx 1.2$).

After the starburst event triggered by the first passage of the last major merger ($z \approx 1.5$), type-II supernovae first enrich the ISM in $\alpha$-elements (on average $\sim 10 \Myr$ after star formation). When the type-Ia supernovae start producing iron significantly later, \afe decreases in the vicinity of the star forming regions, and it takes time for this material to be launched out of the galaxy, especially if the starburst event is already finished and the SFR returns to a low value. The exact time needed for feedback to propagate is a strong function of the turbulent structure of the ISM \citep{Martizzi2015, Kim2015, Ohlin2019} and depends on the details of the alteration of this structure by galaxy interactions and mergers. Such effects include tidal stripping of some regions of the ISM, star formation at large radii (e.g. in the outer disk and/or in tidal tails), the modification of the vertical structure and even the destruction of the disk. All these processes dramatically modify the porosity of the ISM and thus the ability of supernova blasts to find low-density chimneys of least resistance to propagate away from the star forming sites. 

The difference in \afe between in- and outflows reaches its extremum at $z \approx 1$ ($\approx 8 \Gyr$ ago), and keeps it for a relatively long time ($\sim 1 \Gyr$). This corresponds to the outflow-dominated epoch during a period of enhanced star formation activity, after the final coalescence of the last major merger, and the onset of the phase of rarefied interactions. This maximum difference is mainly linked to a decrease in the \afe content of outflows, due to a drop of the $\alpha$ abundance of this gas. We note that this epoch shortly follows the end of formation in the high-\afe sequence (\fig{age}).

At all epochs, the outflows are more metal-rich than the inflows (bottom panel of \fig{outflows}). This is natural at late stages when the accretion is dominated by cold and metal-poor medium \citep{Somerville2015}, and the outflows carry enriched gas. However, we note that the difference in \feh between in- and outflows ($\sim 0.5 \dex$) is already in place during the merger-dominated growth phase (when accretion includes material enriched within the progenitor galaxies), and remains roughly constant through cosmic time. This is likely due to the ejection by mergers of metal-rich material in tidal debris, and galactic winds launched by starburst-triggered feedback. Furthermore, the relatively massive galaxies merging with \vintergatan almost always have a lower SFR and therefore have produced less metals before they merge. In addition, their lower masses suggest a less efficient retention of the enriched winds \citep{Maclow1999}, such that they bring in more metal-poor material (stars and gas) than what is found in \vintergatan at the same epoch, but significantly richer than the diffuse intergalactic medium. In the end, merger events induce both the inflow of richer-than-average material, and the outflow of even richer gas (from the tidal and feedback-driven ejection of part of the ISM).

As a consequence of this complex and multi-scale interplay, we cannot establish a one-to-one causal relation between the starburst events and the periods of low \afe outflows. We note however that such drops are seen after all major mergers, with different amplitudes and durations, thus compatible with an significant influence of the disturbing effects of galaxy encounters. In addition, the peak of star formation seen at $z \approx 1.8$ , which is not directly associated with a major merger event, does not correspond to a significant drop in the \afe content of outflows. This further emphasizes the role of interactions in reshaping the ISM and altering the propagation of enriched material, and thus the necessity to numerically resolve these details, as discussed in \sect{thickness}.

\subsection{Properties of the in situ and accreted populations}
\label{sec:isac}

\subsubsection{Little chemical overlap of the in situ and accreted components}
\label{sec:chem}

\begin{figure}
\centering
\includegraphics{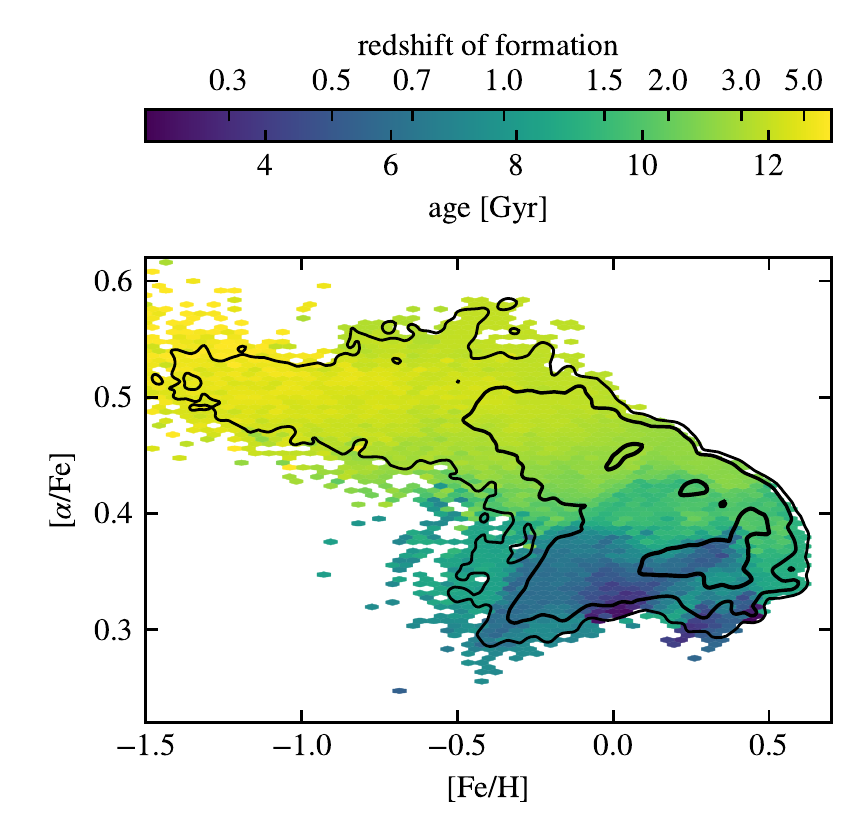}
\caption{Average stellar age in the abundance plane. The black contours indicate the 0.1\%, 1\% and 10\% levels of the total population.}
\label{fig:age}
\end{figure}

\fig{age} shows the average stellar ages in the abundance plane. This was first presented and analyzed in \citetalias{Agertz2020} (their figure 16) and we refer the reader to that paper for a detailed discussion. Here, we simply recall that the high-\afe sequence comprises old stars ($\gtrsim 9 \Gyr$), that the youngest stars are not the most metal-rich, and that stars $\approx 8 \mh 10 \Gyr$ old are found both at the high- and low-metallicity ends of the low-\afe sequence. All these properties match observational data \citep{Bensby2014, Ness2016, Feuillet2019}.

\begin{figure}
\centering
\includegraphics{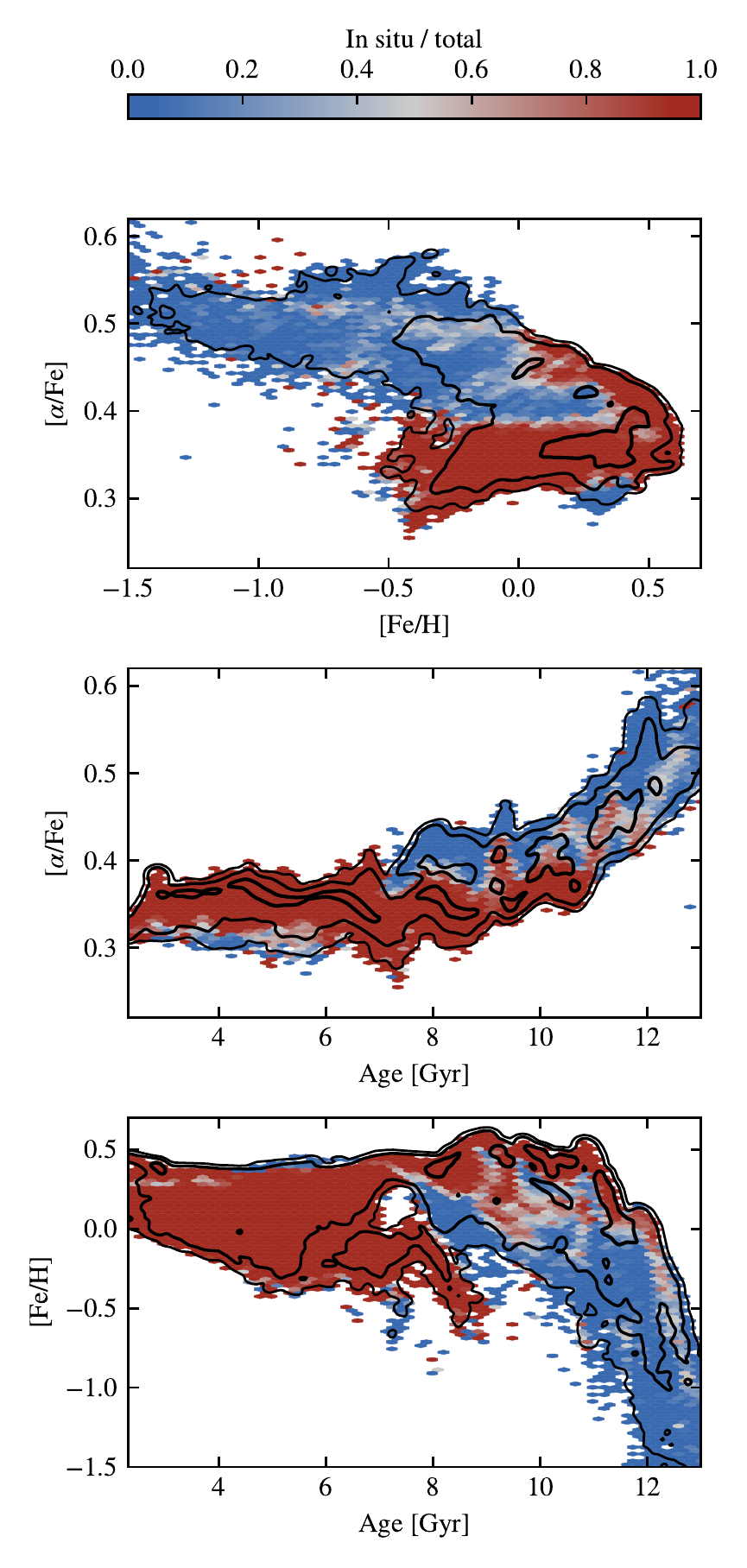}
\caption{Fraction of in situ stars with respect to total (in a grid of $50 \times 50$ hexagonal bins over the ranges shown). The black contours indicate the 0.1\%, 1\% and 10\% levels of the total population. The in situ and accreted populations show little overlap in these planes, with the exception of high \afe stars older than $\approx 10 \Gyr$, i.e. when \vintergatan still resembled most of its neighbor galaxies and thus shared a comparable chemical composition. Although most of the diversity of chemical compositions in the high-\afe sequence is spanned by accreted stars, it does \emph{not} mean that most of the stars in this sequence are of accreted origin: the in situ population covers a narrower variety of abundances, but counts more stars.}
\label{fig:isfrac_contours}
\end{figure}

\fig{isfrac_contours} shows the locations of the in situ and accreted stellar populations, and their overlap in the chemical abundance planes. The different evolutions of the various progenitor galaxies due to their different masses, star formation histories, formation and accretion epochs, lead to little overlap between the two populations. This hints that, up to observational errors, they could be told apart without resorting to additional properties (like other abundances or kinematics), e.g. to retrace the assembly history of the Milky Way (see \sect{linking}). We note however that there is no gap in the abundances between the two populations, which could make this exercise difficult. The effect of observational uncertainties on these distributions is evoked in \sect{detect}.

In the first Gyr of evolution, the majority of stars are accreted and this fraction decreases with time (\fig{sfr}), and thus (at this epoch) with increasing metallicity. As a result, about half of the stars in the high-\afe sequence have an accreted origin, and cover a much wider diversity of chemical compositions than the in situ stars. However, stars in the low-\afe branch are almost exclusively formed in situ. Only a few regions show an overlap of the in situ and accreted populations, mostly spanned by old stars ($\gtrsim 10 \Gyr$, $z \gtrsim 1.8$) in the high-\afe sequence. At a given age, they are found at the high-\afe limit of the in situ population (middle-panel of \fig{isfrac_contours}). 

The striking almost complete absence of overlap of the in situ and accreted populations in the \feh-\afe plane and the fact that, when present, this overlap is most important at the earliest epochs reflect that the assembly history of \vintergatan strongly differs from that of its companion galaxies. Instead, a stronger overlap of the two populations would have implied that the accreted galaxies followed a comparable enrichment history to the main galaxy, i.e. they would have a comparable star formation history (production of metals) and growth (retention of feedback ejecta), and thus a comparable mass evolution. For the same reasons, an evolution involving a very late merger with a mass ratio close to unity would have led to more overlap of the in situ and accreted populations. Therefore, the absence of overlap we note here is a direct consequence of the rapid ($z \lesssim 1.8$, $\lesssim 10 \Gyr$) mass-dominance of \vintergatan over any other member of its group, or in other words, of the absence of another massive galaxy. Recall that the last major merger has a mass ratio of only 1:3 and that the 1:1 mergers occur early, at $z\approx 5.7$ and 2.7, involving low-mass galaxies ($M_\star \sim 10^{7\mh 9} \msun$) and thus not contributing much to the final mass budget of \vintergatan (\fig{sfr}). By estimating the in situ or accreted origin of stars (e.g. observationally from kinematics), the degree of overlap of the two populations in the \feh-\afe plane is thus an indicator of the predominance level of the galaxy over its neighbors at different stages of its evolution.

\subsubsection{Chemical bimodalities}
\label{sec:bimodalities}

\begin{figure}
\centering
\includegraphics{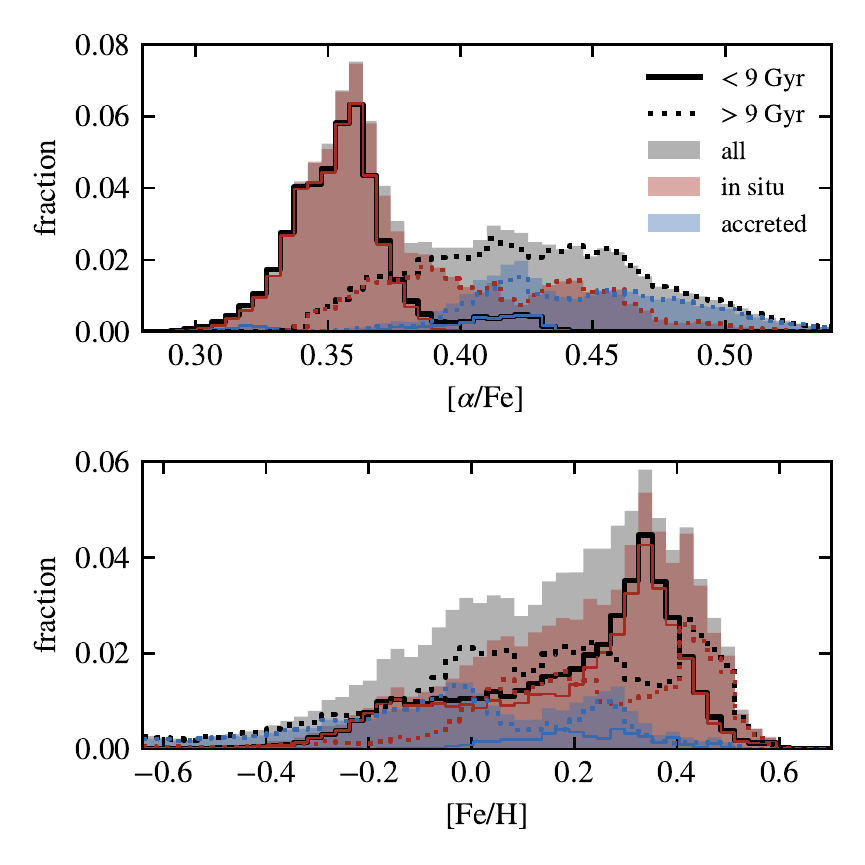}
\caption{Distributions of \afe and \feh for the in situ and accreted populations. Dotted and dashed lines split the populations into the contributions in the merger-dominated ($> 9 \Gyr$) and quiescent phases ($<9 \Gyr$).}
\label{fig:isac_histo}
\end{figure}

\fig{isac_histo} shows the individual distributions of \afe and \feh for the in situ and accreted populations, separating the stars formed during the merger-dominated phase of the assembly (approximately $> 9 \Gyr$) from those formed later. As already hinted in \fig{isfrac_contours}, the little overlap of the in situ and accreted population becomes less evident in these projected distributions. This shows the importance of combining more than one tracer when identifying the origins of sub-populations.

Apart from the tail made of the oldest stars, the high-\afe regime appears as a wide plateau ($0.39 < \afe < 0.47$, corresponding to the high-\afe branch), with approximately a half-half contribution of in situ and accreted stars, and mostly formed early. The accreted distribution reaches it maximum at $\afe \approx 0.42$ with the contribution of the last major merger (see \sect{lastmm}). The distribution of accreted stars steadily declines as \afe decreases, as the result of the end of the merger-dominated phase. As noticed above, the low-\afe sequence is narrow ($< 0.1 \dex$) and almost exclusively made of in situ stars, formed after the last major merger ($\lesssim 9 \Gyr$, $z \lesssim 1.3$). 

In terms of \feh, the global distribution has a bimodal behavior, with the metal-rich peak centered on $\feh = 0.35$ and the metal-poor distribution showing a long tail and peaking at $\feh = 0$.  The in situ population mostly contributes to the high metallicity peak, although its distribution yields a tail at low \feh in which the accreted population dominates the mass budget, due to the contribution of many satellite galaxies.

At early times, \vintergatan is still a low-mass galaxy with a low SFR, and thus a slow chemical enrichment, further slowed down by the loss of most of its enriched gas in outflows (due to the shallow galactic gravitational potential at that time, \citealt{Tonini2013, Renaud2017}). The same situation occurs in the satellite galaxies accreted early. As a consequence, the distribution of the old, metal-poor, accreted component resembles that of its coeval in situ counterpart. Later, the growth of \vintergatan accelerates (as the most massive galaxy in its environment, e.g. \citealt{Moster2013}), which allows for a self-enrichment earlier and faster than in any other companion galaxy. Consequently, the old in situ population contains higher-metallicity stars than the old accreted population. We note that metal-poor stars can also form in situ at late stages ($\lesssim 8\mh 9 \Gyr$) in the outer disk, as discussed in \sect{lastmm} and \citet{Renaud2020b} (hereafter \citetalias{Renaud2020b}).

The increase of the in situ fraction with time (\fig{sfr}) induces a skewed distribution in \feh. The concomitant decrease of the merger rate lowers the relative contribution of the accreted stars in the metal-rich peak ($\feh \approx 0.2$) such that, apart from the contribution from the last major merger at $\feh \approx 0.25$, the accreted population peaks at a lower metallicity than the in situ. The superposition of these two distinct peaks leads to the bimodality of the global \feh distribution. Therefore, the bimodality in the metallicity distribution directly results from the transition from the merger-dominated growth to a slower, secular growth of the galaxy. Another galaxy with either a different growth rate (i.e. a different self-enrichment rate) and/or a different merger history would thus yield a different bimodality, or even the absence of a bimodality (e.g. in a case of a longer or still on-going merger-dominated phase, as observed for instance in M~31, \citealt{Gregersen2015}).

The distributions of accreted and in situ stars as function of radius and height in the simulated galaxy are provided in \app{position}. Without discussing them in detail, we note that \fig{isac_position} is in qualitative agreement with the results of \citet{Hayden2015} from APOGEE data (see also \citetalias{Agertz2020}).

\subsubsection{Kinematics}

On top of the contribution of their own stars by accretion, incoming satellite galaxies induce tides which excite the orbits of the in situ population \citep{Toth1992, DOnghia2016}, and thus alters their kinematics \citep{Qu2011, Haywood2018}. For instance, the last major merger induces an increase of $\sim 3 \kms$ (i.e. about 5\%) in the velocity dispersion of the in situ stars for a few $100 \Myr$.

\begin{figure}
\centering
\includegraphics{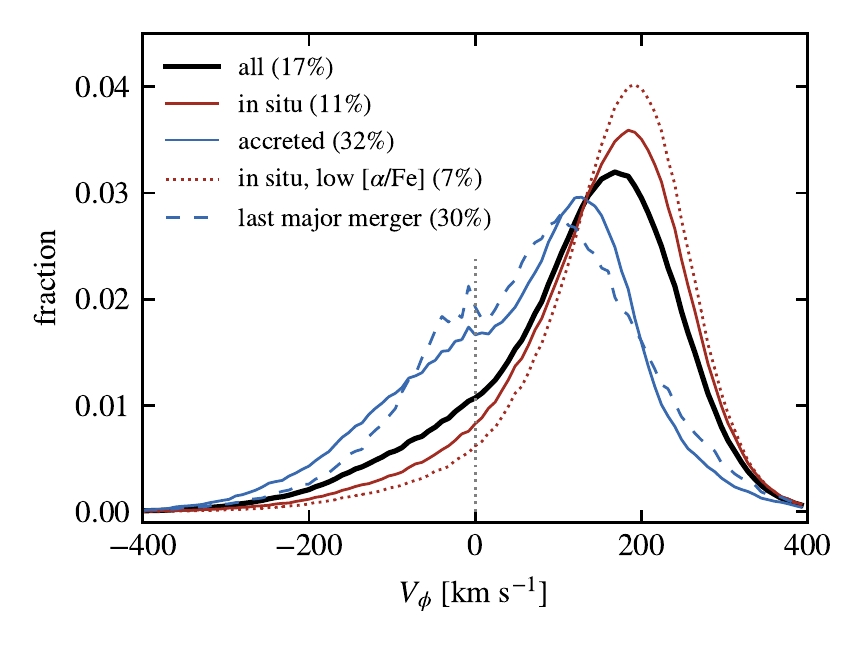}
\includegraphics{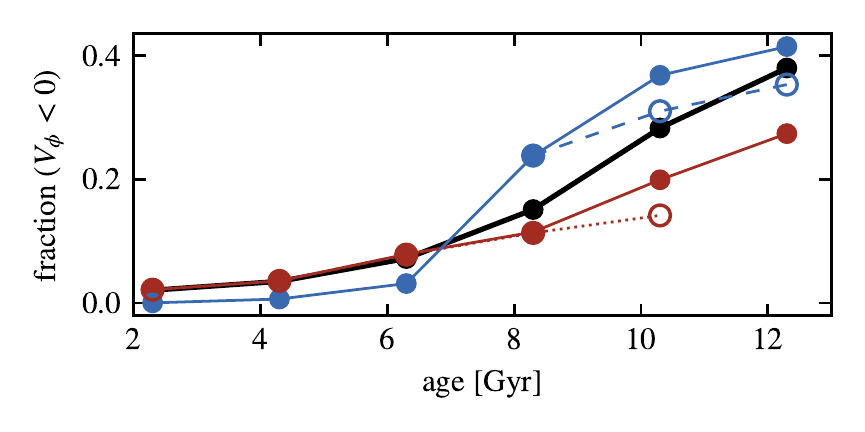}
\caption{Top: normalized distributions of tangential velocity components in the plane of the disk (with a positive velocity in the direction of the disk' rotation) for the entire simulated galaxy (black), the in situ stars (solid red), in situ stars in the low-\afe sequence ($\afe < 0.39$, dotted red), the accreted stars (solid blue), and stars from the last major merger (dashed blue). The numbers in the legend indicate the percentage of retrograde stars for each population. Bottom: fraction of retrograde orbits of each sub-population as a function of stellar age.}
\label{fig:counter}
\end{figure}

\fig{counter} shows the distribution of tangential velocities for the in situ and accreted stars. As expected from the complex merger history and an a priori misalignment of the satellites, a larger fraction of the accreted stars than the in situ component are found with retrograde motions. However, this fraction remains lower than 50\% (as one would expect from a random distribution of the merger parameters), which reflects two aspects: first is the non-isotropic large scale structure of the cosmic web which favors prograde encounters (i.e. aligned with structures that give the main disk its global angular momentum, \citealt{Dubois2014}), and second is the progressive alignment of the accreted material with the disk due to gravitational torques \citep{Codis2012} and dynamical friction \citep{Read2008}.

Stars from the last major merger follow a comparable trend as the overall accreted population, but with a notable secondary peak of the distribution at $V_\phi=0$, corresponding to radial orbits. Such radial orbits are compatible with the observed signature of the Gaia-Enceladus-Sausage galaxy (\citealt{Belokurov2018, Helmi2018}). In addition, the epoch of the last major merger and the mass of the progenitor galaxy in our simulation (interaction between $9.5 \Gyr$ and $8.7 \Gyr$ and a virial mass of $7.3 \e{10} \Msun$) are remarkably similar to that derived from observational data of the Gaia-Enceladus-Sausage (merger $9.5 \Gyr$ ago, and a virial mass $> 10^{10} \Msun$, \citealt{Belokurov2020}). The detection of the sausage-shaped feature in velocity space of the Milky Way has been done at low metallicity ($-3 < \feh < -1$), which is very sparsely populated in \vintergatan: an analysis of these stars would strongly suffer from low-number statistics. However, we argue that a galaxy as massive as the Gaia-Enceladus-Sausage would likely have enriched in a comparable manner as the Milky Way at the same epoch, and thus should also comprise stars at higher metallicities (as found here and in other simulations, e.g. \citealt{Grand2020}). Therefore, it is possible that, like in our last major merger, the Gaia-Enceladus-Sausage also brought a component at high metallicity and non-zero tangential velocities in addition to that detected so far. This component would strongly overlap with the in situ population in a number of parameters, making its identification very involved. This additional, yet to be detected component would then call for a revision of the mass of the Gaia-Enceladus-Sausage. Compared to the existing data however, our last major merger has an overall too high prograde velocity (with its peak at $\approx 100 \kms$). The origin of this component and its sensitivity to the properties of the merger will be explored in a forthcoming paper.

The bottom panel of \fig{counter} shows that, for all ages, retrograde motions always concern a minority of stars. However, this fraction monotonically increases with age, especially in the accreted component. The majority of in situ stars on retrograde orbits (11\% of the in situ stars and 17\% of the total simulated galaxy) form before the disk itself ($z\gtrsim 4$, see \citetalias{Agertz2020}) and thus do not inherit disk-like kinematics. \citet{JeanBaptiste2017} propose that in situ stars could evolve to retrograde motions after being dynamically heated by mergers, again advocating for a larger fraction of retrograde stars formed before the last major merger. 

\begin{figure}
\centering
\includegraphics{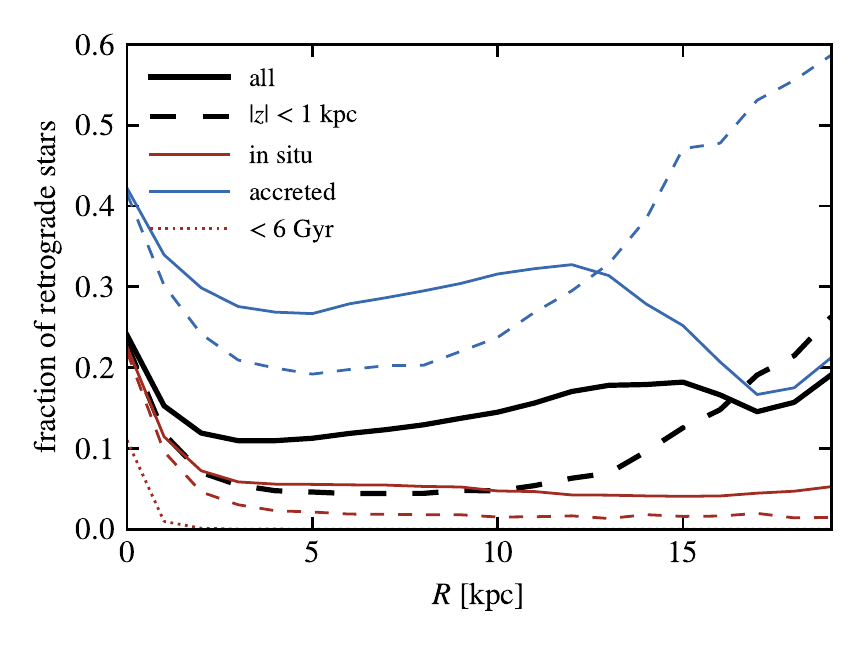}
\caption{Radial distribution of all stars on retrograde orbits (solid lines), and for stars within 1 kpc from the mid-plane of the disk (dashed lines). The red dotted line corresponds to stars younger than 6 Gyr, i.e. formed after the alignment of the inner and outer disks (see text and \citetalias{Renaud2020b}), only shown for the in situ component due to scarcity of young accreted stars. Apart from the contribution of recent mergers in the outer disk ($\gtrsim 10 \kpc$), most of the retrograde stars are found in the inner galaxy, in particular in the innermost kpc for the in situ component.}
\label{fig:retro_radial}
\end{figure}

\fig{retro_radial} shows the fraction of stars on retrograde orbits as a function of radius. Almost all the in situ stars on retrograde orbits are found in the central-most region of the simulated galaxy, i.e. in the region dominated by the dynamics of the bulge and where disk kinematics have a weaker imprint on the ISM and the newly formed stars. As expected, this trend is even stronger when considering the young component only: 94\% of in situ stars younger than $6 \Gyr$ (i.e. formed after the alignment of the inner and outer disks, see figure 8 of \citetalias{Agertz2020} and \citetalias{Renaud2020b}) are found within the innermost kpc, and 100\% are within $3 \kpc$. The older in situ retrograde component extends further, as a signature of star formation in a turbulent and thick disk subjected to repeated tidal disruptions \citep{DOnghia2016}.

Apart from the innermost kpc, the fraction of in situ stars on retrograde orbits is approximately constant with radius ($3\mh 5 \%$). Conversely, retrograde accreted stars are strongly radially concentrated. In addition to a central peak, their fraction also increases at large radius but only at high heights ($\gtrsim 12 \kpc$, $|z| \gtrsim 1 \kpc$), indicating late accretion (via minor mergers and tidal debris) and a long timescale needed to erase kinematic features in the shallow potential of the off-plane outer galaxy.

\subsection{Major mergers, starbursts, and the onset of the low-\afe sequence}
\label{sec:lastmm}

\subsubsection{Contributions of major mergers to the high-\afe sequence}

\begin{figure*}
\centering
\includegraphics{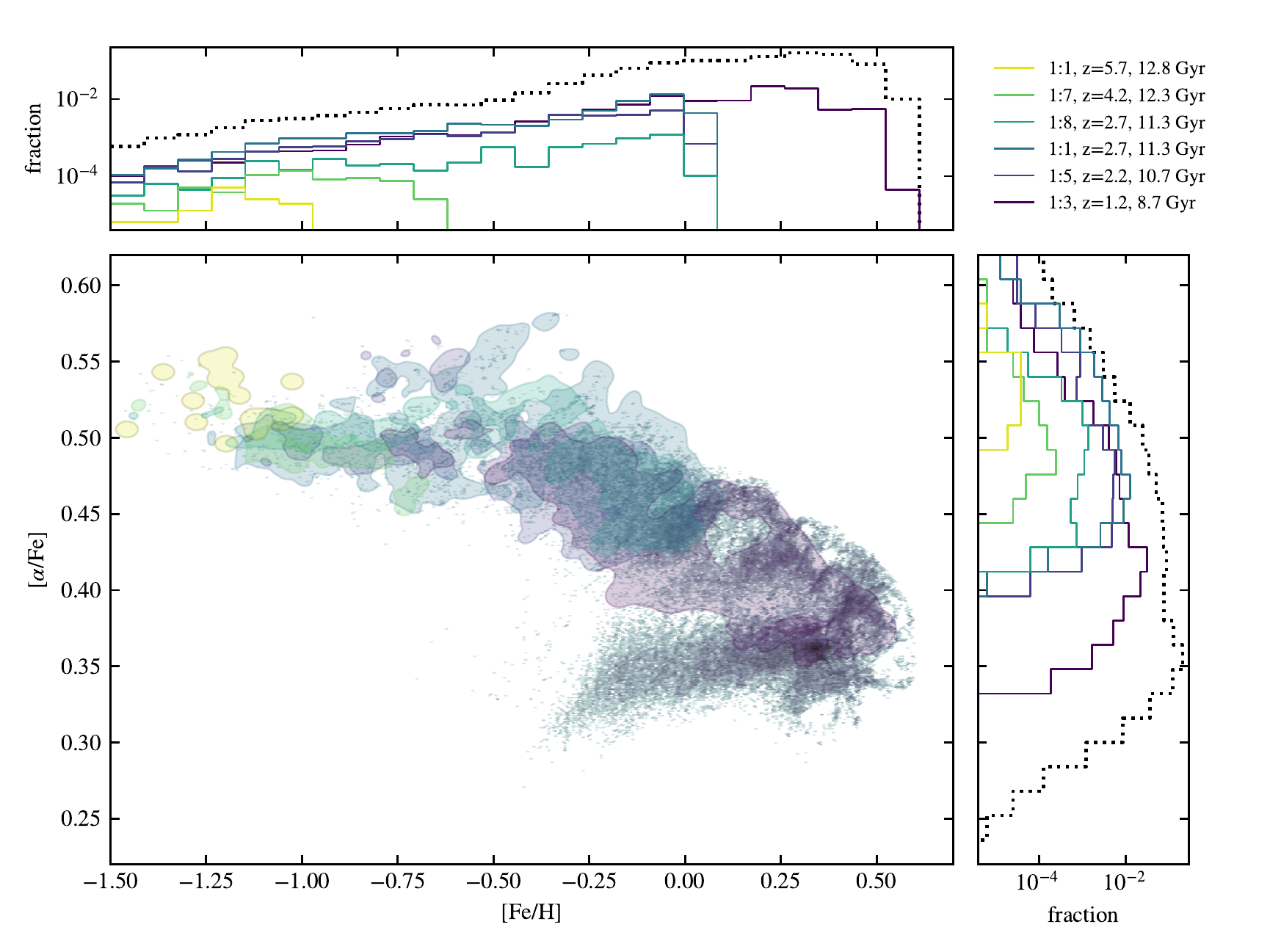}
\caption{Contributions of the 6 major mergers to the final chemical composition. The mass ratio and epoch of coalescence of the mergers are indicated in the top-right corner. The contours are smoothed for clarity and mark the locations of the 1\% level of the maximum of each galaxy. The gray scale background and the gray dotted lines in the histograms show the distribution of all \vintergatan's stars. The histograms show the projected distributions, normalized to the total number of stars in \vintergatan. All major mergers reach the low (resp. high) end of the \feh (resp. \afe) distributions, but their extent toward the other end varies roughly (but not exactly) with the redshift of their accretion. The low-\afe branch counts some stars from the last major merger, but the very large majority of the material accreted during the major mergers only spans the high-\afe sequence.}
\label{fig:mm}
\end{figure*}

\fig{mm} shows the contributions of all the major mergers (i.e. with a mass ratio higher than 1:10) in the abundance plane. As discussed in \sect{isac}, the similar masses of these galaxies with \vintergatan at the epoch of their interaction (by definition) implies similar chemical contents, yet with non negligible differences owing to different star formation histories, and non exactly equal masses. 

The major mergers populate almost exclusively the high-\afe branch, and only the late formation in the last major merger reaches the low-\afe regime, at high metallicities and at the very end of the merger-dominated phase ($\approx 8\mh 9 \Gyr$). At a given metallicity, the stars from a major merger span a significant fraction of the width of the \afe sequence, but it is the superposition of several distinct galaxies (including minor mergers) that makes the full width of the \afe branch (see also \fig{isfrac_contours} which highlights the relatively narrow spread of the in situ population at high \afe). 

From their simulation of a galaxy in isolation, \citet{Clarke2019} proposed that high \afe would be reached when stars form at high surface density of SFR (as suggested by chemical evolution models, e.g. \citealt{Matteucci2012}). In their setup without interactions and mergers, such physical conditions are found when the disk develops large-scale instabilities in the form of massive clumps (as observed in gas-rich disks, e.g. \citealt{Wuyts2012}). Therefore, this requires the pre-existence of a non-clumpy disk, thus forming stars at low \afe. As a consequence, in their scenario, the high-\afe sequence gets populated once the clump instabilities appear, i.e. from the low-\afe branch. Despite remarkable matches of the chemical properties of the simulation of \citet{Clarke2019} with APOGEE data (see also \citealt{Beraldo2020, Amarante2020}), this mode of formation is in tension with observations of the high-\afe sequence being overall older than the low-\afe one \citep{Bensby2014, Silva2018, Feuillet2019}. Nevertheless, a similar mechanism would account for the observed age distribution if the episodes of intense star formation were to be ignited by mergers \emph{before} the formation of the low-\afe sequence, i.e. directly from the high-\afe sequence\footnote{This would however fail at explaining the bimodality in \afe, calling for another argument, as discussed later.} rather than by massive clump instabilities, as shown below. This problem is solved in the scenario from \citet{Khoperskov2020}, in which the intrinsic evolution of the star formation regime explains the transition between the two sequences. Metallicity gradients and the radial dependence of the star formation timescale then account for the complex build-up of the low-\afe sequence. Therefore, despite differences with observational data, simulations of isolated galaxies can successively serve as proofs of concept for an intrinsic, non-cosmological origin of the chemical bimodality. It is thus important to establish to which extent external effects, in particular mergers, alter, amplify, reduce, accelerate or delay the outcomes of internal evolution. In \sect{scenario}, we argue that internal and external effects work together to allow for the transition between the thick, kinematically hot, high-\afe disk, to the thin, colder, and younger low-\afe sequence.

\subsubsection{The effects of starbursts}
\label{sec:starbursts}

The absence of a \afe bimodality in the major merger progenitors indicates that such a bimodality appears at a relatively late stage of galaxy evolution (hence only in the \vintergatan galaxy), either after the merger-dominated growth, or once the galaxy has reached a sufficient mass. To test these hypotheses, we now examine the relation between the chemical content and the starburst activity induced by major mergers.

A starburst is characterized by a \emph{fast} formation of stars, i.e. a short depletion time (defined as the ratio of gas mass to SFR). This is not to be confused with the formation of many stars, as traced for instance by a high SFR, which could correspond to an intense, yet steady and not bursty, star formation activity (see \citealt{Renaud2019b}). A starburst episode triggers the production of core-collapse (type-II) SNe at a suddenly enhanced rate with respect to the usual type-Ia rate from previous star forming events \citep{Ruchti2010}. Therefore, during the time delay between the onset of type-II and type-Ia SNe, the $\alpha$ content of the interstellar medium surrounding a starburst event rapidly increases, without the immediate outbreak of iron. As a consequence, with little change in metallicity, the \afe ratio rapidly increases during a starburst episode. We stress that the increase is relative to the previous level: a starburst does not necessarily lead to high \afe, but to a \emph{temporary enhancement} of \afe. Assuming that the physical conditions for star formation are still present in the enriched gas, new stars are formed with this boosted \afe. When the progenitors of type-Ia SNe formed during the starburst finally go off, they increase the iron content of the medium and, again assuming that star formation continues (which is less likely than before because of the longer time delay), they bring the \afe ratio of the stars to lower values than the pre-starburst levels.

This situation is well reproduced in chemical evolution models (see e.g. \citealt{Johnson2020, Thorsbro2020}). In such models, an ad hoc starburst event is described by instantaneously adding (pristine) gas to the galaxy, which induces a rapid dilution of the metallicity, quickly followed by the effect of the starburst described above. The resulting signature is a clockwise loop-shape in the \feh-\afe plane (e.g. the top-center panel of figure 1 in \citealt{Johnson2020}). In our simulation, however, the infall of intergalactic gas onto the galaxy is smooth and fuels the galactic halo and corona rather than the star forming regions directly, thus not leading to any starburst activities. Only the major galaxy interactions and mergers can trigger a rapid boost of star formation, by compressing and/or shocking the gas already present in the galaxies \citep{Renaud2019b}. This implies that the starbursting medium has a comparable chemical composition as the bulk of the galaxy progenitors, and thus that the initial drop in metallicity seen in chemical evolution models does not occur, only the temporary enhancement of \afe. (The loops are replaced with hooks.) This could correspond qualitatively to the features identified by \citet[their figure 2]{Zasowski2019} in the inner Galaxy, and by \citet{Thorsbro2020} in several stars of the nuclear star cluster. We note that a similar hook-shaped feature has been detected in the Large and Small Magellanic Clouds by \citet{Nidever2020}. If these features are indeed of the same nature as to those found in \vintergatan, this would advocate for a recent galactic-wide starburst in the Magellanic Clouds, likely triggered by an interaction between the pair and/or between them and the Milky Way.

\begin{figure}
\centering
\includegraphics{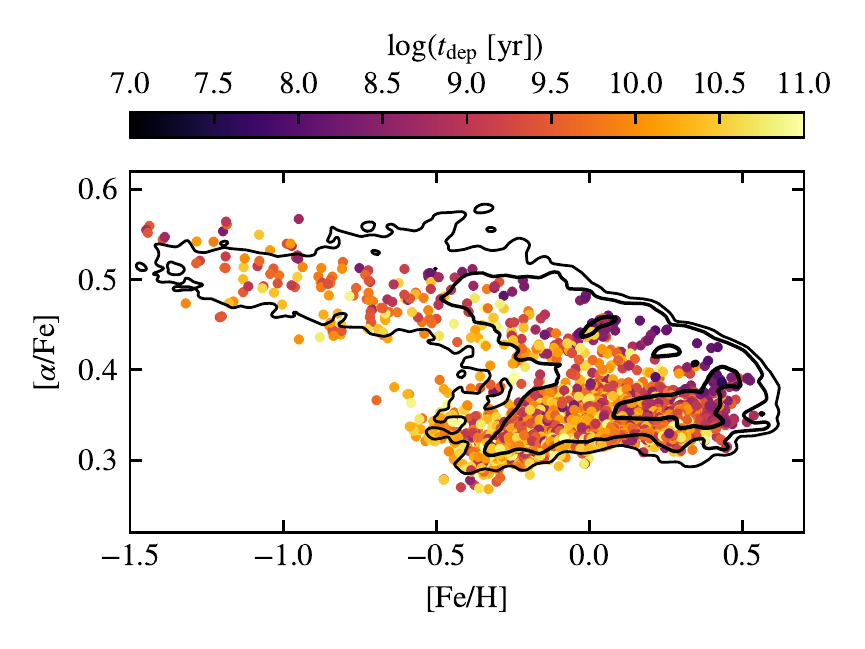}
\caption{Distribution of gas depletion times in regions of $1\kpc \times 1\kpc$ in the simulated galaxy, stacked over the entire simulated period. The abundances are that of the stars younger than $10 \Myr$ (see text for details and caveat). The shortest depletion times are found on the upper side of the high-\afe sequence, at all metallicities and all epochs. The black contours indicate the 0.1\%, 1\% and 10\% levels of the total population, to guide the eye.}
\label{fig:tdep}
\end{figure}

\fig{tdep} shows the distribution of gas depletion times in the abundance plane. For each of our snapshots, the depletion time \tdep is computed in regions of $1\kpc \times 1 \kpc$ in the simulated galaxy, as the ratio of the surface density of gas and the surface density of SFR measured using stars younger than $10 \Myr$.\footnote{Because the average time difference between our snapshots is longer than $10 \Myr$, not all stars are represented here, which explains that many star formation events are not captured by this approach, and why some areas of this plot are not populated.} Here, the abundances are computed as the average of that of these young stars in each region. Finally, the results from all the snapshots are stacked into this figure. Note that our measure of the depletion time is a direct proxy for the \emph{current} star formation activity, while the abundances reflect enrichment by \emph{former} generations of stars. Therefore, our comparison between depletion time and chemical enrichment is not direct. However, we argue that the starburst episodes last generally longer (at kpc scale) than the time difference introduced by our method, such that our conclusions are not affected by this artifact.

The last major interaction (first passage and coalescence, $\approx 8.7\mh 9.5 \Gyr$, $z\approx 1.2\mh 1.3$) induces the last significant star formation boost in \vintergatan (\fig{sfr}). The stars formed during this event are found at the high-\afe and high-\feh end of the distribution. As visible in \fig{tdep}, they are exclusively associated with short depletion times, further confirming the starburst nature of this event. As expected from the arguments above, the burst triggers an enhancement of the \afe content with respect to the typical chemical content of young stars immediately before the interaction, explaining the position of these stars at the upper end of the \afe distribution at this epoch. \fig{isfrac_contours} further indicates that these stars are formed in situ.

With the mass growth of galaxies in the merger-dominated phase ($\gtrsim 8\mh 9 \Gyr$), significant starbursts (in term of stellar mass formed and reduction of the depletion time) start to concern only the most massive galaxies (in our case, \vintergatan and the progenitor of the last major merger), while lower mass galaxies rather undergo tidal disruption, ram pressure stripping and/or quenching during interactions \citep{Tinker2013, Simpson2018}. Furthermore, the intensity of starbursts (as high SFRs and short depletion times) increases with decreasing galaxy separation \citep{Scudder2012}. This implies that most of the stars formed during a starburst event are likely to form when the two progenitor galaxies are on the verge of merging or have already merged, which means that most of the formation is considered to happen in situ in the resulting galaxy. Therefore, in the early Universe when all galaxies still have similar masses, most galaxies can host an enhancement of their \afe content triggered by the starburst associated with a major merger, which is thus seen in both the in situ and accreted content of the final galaxy (\fig[s]{isfrac_contours} and \ref{fig:mm}). However with time, the main galaxy starts to dominate the mass budget of its environment, and the population of stars with \afe enhanced by starbursts is eventually mostly found in the in situ component, as noticed in the case of the last major merger.

\subsubsection{Transition to the low-\afe sequence}

Gravitational torques and tides associated with the interactions alter the structure of disks, enhancing the turbulence and the density contrast between dense gas clumps and their surroundings. The sharp density gradient around the clumps (about 3 orders of magnitude difference in density between a clump and its edge, i.e. over $\sim 500 \pc$, at $z\approx 1.5\mh 2$) means that very low densities exist in the immediate vicinity of the star forming regions. This effectively increases the porosity of the ISM, which favors the coupling of feedback to large scales ($\gtrsim 1 \kpc$), and the launching of strong galactic winds (\fig{outflows}). This effect would be further enhanced if considering feedback from runaway stars \citep{Andersson2020}. The enriched material propagates to large distances (up to several $10 \kpc$ from the galaxy) where it mixes with lower metallicity media. This explains why the star forming regions become less efficient at self-enriching, and why the metallicity of the inflows starts to saturate. Thus, \feh reaches its maximum shortly after this epoch ($\approx 10 \Gyr$, $z=1.8$). The interaction of the outflowing material with the hot corona, including the compression in the shock fronts of bubbles \citep{Hobbs2015}, favors its cooling and condensation onto the disk \citep{Fraternali2013}. Later ($\lesssim 7 \Gyr$, $z \lesssim 0.8$), this gas, now diluted with lower metallicity medium from the halo, eventually fuels the star forming disk. Together with chemical mixing with the low-\feh content of the outer disk (\citetalias{Agertz2020}), this participates in the inside-out growth of the disk \citep{Bird2013}, and allows for the formation of stars in the low-\afe branch at metallicities not higher than that reached during the last major merger. Despite a less clumpy and turbulent ISM (and thus a less efficient launch of outflows), the galaxy does not self-enrich anymore, likely because of the significantly lower SFR after the merger phase ($\lesssim 15 \Msunyr$).

After the last major merger, star formation in \vintergatan enters the non-starburst phase of long depletion times, and thus lacks rapid enhancements in \afe. This favors the transition from the high- to the low-\afe sequences. We note that relatively low values of \afe ($\lesssim 0.39$) are reached as early as $10 \Gyr$ ago at high metallicity (\fig{isfrac_contours}), likely because of a significant decrease of the SFR (and elongation of the depletion time) in between interactions at this epoch. However this population rather constitutes the continuity of the high-\afe sequence than the onset of a new, distinct formation mode. These stars are found to have intermediate vertical velocity dispersions and thickness, significantly lower than the bulk of the high-\afe sequence, but also higher than most of the low-\afe population (figure 16 of \citetalias{Agertz2020}). They effectively mark the transition between the geometrical thick and thin disks, where the high- and low-\afe sequences overlap, in agreement with the observations of \citet{Bovy2016}. The metal-rich part of the low-\afe sequence is thus the superposition of the late build-up of the high-\afe branch (formed $9\mh 11 \Gyr$ ago), and the onset of formation in the thin disk after the cessation of the merger activity. As such, this region of the \afe-\feh plane contains the stars for which the chemical composition correlates the least with age.

After the metallicity has already reached its maximum (\fig{isfrac_contours}), the overall decrease of \afe cannot be explained by a rise of \feh (e.g. due to old type-Ia SNe), but rather by a decrease of the production of $\alpha$ due to quieter SFRs and longer depletion times. Furthermore, the end of tidal influence by interactions and mergers lowers the orbital excitation to high altitude above/below the plane of the disk, as well as the level of turbulence in the disk \citep{Renaud2014b}. This helps the galaxy to dynamically cool down, marking the end of star formation in the turbulence-supported and tidally-excited thick disk. 

Concurrently, the first pericenter passage of the last major merger triggers the non-starburst formation of stars at the low-metallicity end of the low-\afe branch in the outer disk, as discussed in detail in \citetalias{Renaud2020b} and summarized in \sect{features}. The coeval formation of low- and high-\feh stars in the low-\afe sequence at $\approx 6\mh 9 \Gyr$, in line with observation data (e.g. \citealt{Feuillet2019}), indicates the co-existence of two distinct formation channels in the low-\afe sequence.

All these events take place within about 1 Gyr around the epoch of the last major merger, and participate in the transition between the two \afe regimes, and from the thick to thin disk. This transition results from the interplay of external effects, as the end of the merger bombardment, and intrinsic evolution, as the end of the clumpy phase and the onset of the secular evolution in an extended thin disk. As such, it reconciles the scenarios derived from simulations of isolated disks in which the transition results from the intrinsic evolution of the star formation and enrichment regime \citep[e.g.][]{Clarke2019, Khoperskov2020} and the models invoking different phases of star formation due to discrete accretion events \citep[e.g.][]{Grand2018, Spitoni2019, Buck2020, Lian2020}. With the tilting disk scenario, \citetalias{Renaud2020b} shows why both external and internal transitions are likely synchronized, due to the inside-out formation of the extended disk initiated by the last major merger.

\subsection{Linking chemical signatures to their physical origins}
\label{sec:linking}

\subsubsection{Imprint of the main phases of galaxy evolution}

\begin{figure}
\centering
\includegraphics{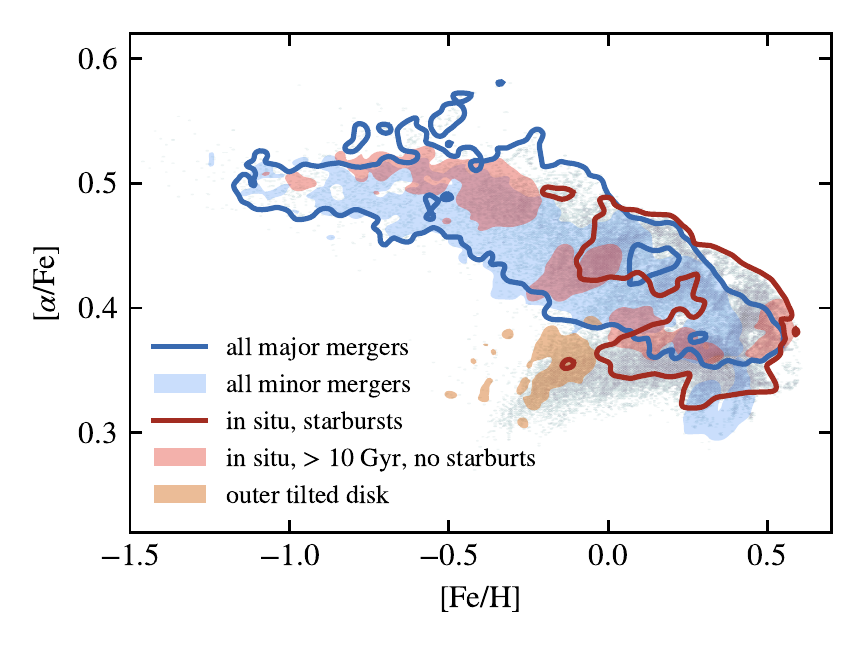}
\caption{Identification of sub-populations associated with the main phases of galaxy formation. The contours are smoothed for clarity and mark the locations of the 1\% level of the maximum of each population. The gray scale background shows the distribution of all the stars in \vintergatan. The ``starburst'' criterion identifies stars formed during the merger-induced peaks of star formation (arbitrary set at $> 15 \Msunyr$ and for ages $> 7\Gyr$, see \fig{sfr}). We note that this also includes stars formed during these epochs, but with no starburst physical conditions (i.e. with long depletion time), but that these stars represent a minority of this population. The ``outer tilted disk'' gathers the stars formed at low metallicity in the initially tilted outer disk before it aligns with the inner disk (see below, and \citetalias{Renaud2020b} for details and complementary diagnostics). This population is selected as stars formed between $7.8$ and $9.2 \Gyr$ and with $\feh < 0$.}
\label{fig:selection}
\end{figure}

\fig{selection} shows the locii of stars formed during certain phases of galaxy formation in the \feh-\afe plane. It confirms the four general trends highlighted above. (\emph{i}) The major mergers are almost exclusively found in the high-\afe sequence. (\emph{ii}) At a given metallicity, minor mergers occupy the low-\afe side of this sequence. (\emph{iii}) Conversely, stars formed during the starburst phases are on the high-\afe side of the sequence. Stars in the ``starbursts'' group at low \afe are formed during the episode triggered after the coalescence of the last major merger, but are not associated with particularity short depletion times (\fig{tdep}). These stars are rather the quiescent counter-part of this episode, and thus do not trace starburst-like physical conditions. (\emph{iv}) The early star formation in the outer disk yields a different chemical composition from the other populations (\citetalias{Renaud2020b}).

In the Milky Way, dwarf satellite galaxies and streams like the Large and Small Magellanic clouds, Leo A, Sagitarius and others show a drop of the \afe abundance at a significantly lower metallicity than the Milky Way itself \citep[e.g.][]{Hasselquist2019, Nidever2020}, which has been interpreted as longer depletion times in the low-mass systems \citep{Andrews2017}. Such galaxies are absent of \vintergatan, because all the minor mergers in the simulation contribute to the high-\afe sequence, with comparable (yet not identical) chemical compositions as the main galaxy. Identifying the reason(s) for such a discrepancy with the observations would require to explore alternative merger histories, which we leave for a forthcoming paper. However, we speculate that it could be due to either a dearth of late-infalling satellites (after the last major merger), a lack of resolution in the formation of the low-mass galaxies biasing their formation efficiency, a lack of resolution in the hot halo of \vintergatan in which such galaxies could be polluted during their infall, or a combination of these factors.

Apart from the peculiar locus of the outer disk, there is no unambiguous signature from the other groups which largely overlap in chemical space. The reason for this is the wide diversity of conditions in which these events occur, leading to a large scatter in the resulting abundances, and ultimately to the degeneracy of the link between physical origin and position in chemical space. The same applies when considering abundances and stellar ages. In other words, a given region in the chemical space cannot be unambiguously identified as the result of a precise and unique phase of galaxy formation, despite the moderate overlap of the in situ and accreted populations noted above (\sect{chem}). However, \fig{selection} suggests that some precise features in this plane could result from a unique physical mechanism applied to a narrow range of initial conditions. We explore this possibility in the next section.

\subsubsection{Origins of notable features in the \feh-\afe plane}
\label{sec:features}

\begin{figure}
\centering
\includegraphics{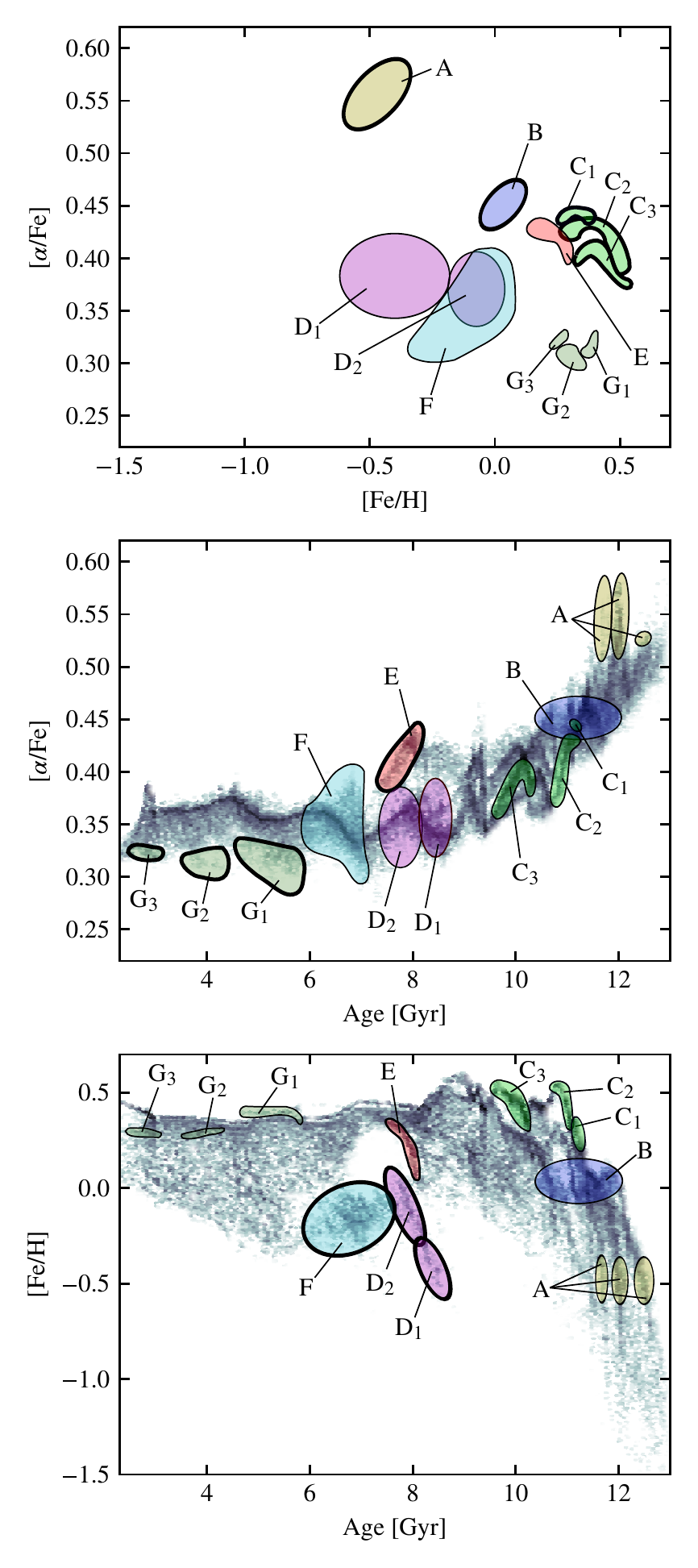}
\caption{Identification of the main features discussed in \sect{features}. A thick border indicates in which panel each feature has originally been selected. The structures are identified by eye, as overdensities of stars in at least one of the panel of this figure, and are ordered chronologically. The gray scale background shows the distribution of all the stars in \vintergatan. Number of other overdensities in these planes do not correspond to clear signatures in other dimensions (e.g. kinematics), which implies that they represent the superposition of several distinct physical origins and thus cannot be traced back to a single, simple event along galaxy evolution.}
\label{fig:features}
\end{figure}

\fig{features} shows our (arbitrary) identification of notable features in the plots of \afe, \feh and age. By examining the properties of the stars from theses selections in a wider parameter space (chemistry, morphology, kinematics, ages, location etc.), we retrace their origins in order to better understand the mechanisms involved and their interplay, as a first attempt to connect generic physical events to their generic signatures in observable space.

The main differences between \vintergatan and the real Galaxy are due to the stochasticity of the merger events, the finite resolution, and the incompleteness and the imperfection of the physical recipes inherent to any simulation. The effect of accounting for observational uncertainties is discussed in \sect{detect}. The analysis presented here could be used as a theoretical framework for the interpretation and the decryption of observational data, and could potentially bring constraints to the input of chemical evolution models.

\emph{Feature A} is found at high \afe and relatively low metallicity. This group encompasses three regions of distinct ages (12.5, 12.0 and 11.7 Gyr) sharing the same chemical composition. The oldest region gathers stars both formed in situ and accreted, but the other two regions, with the highest \afe, are exclusively made of accreted stars (recall \fig{mm}). The mere intrinsic chemical spread of these progenitors creates the illusion of a sequence in the abundance plot. This shows that even well-contrasted and relatively isolated features in the abundance plot can comprise stars of various origins, formed several $100 \kpc$ and $100 \Myr$ apart, due to similarities in the chemical enrichment of different galaxies at high redshift.

\emph{Feature B} spans over 1.5 Gyr in age, at the epoch when the galaxies undergo rapid chemical enrichment. This feature is characterized by a more rapid enrichment in $\alpha$ than in Fe, corresponding to repeated bursts of star formation (\fig{tdep}). It encompasses stars formed both in \vintergatan and in the last major merger (before their first encounter), implying that, just like in feature A, the two galaxies experience a comparable yet still independent evolution at this epoch. As with \fig{selection}, this shows that the stochasticity of the starbursts is smoothed out in chemical space, where the two distinct histories become indistinguishable.

\emph{Features C$_1$, C$_2$} and \emph{C$_3$} are three distinct groups of high metallicity, high \afe in situ stars formed in the starbursts triggered by several major mergers (\fig[s]{isfrac_contours} and \ref{fig:tdep}). These features are hook-shapes structures resulting from rapid enrichment in $\alpha$ during starburst events (comparable to results from chemical evolution models, e.g. \citealt{Grieco2015, Johnson2020}, as discussed in \sect{starbursts}). Because of the diversity of timescales for the onset of type-Ia SNe, the increase of Fe is much slower ($\approx 0.05 \dex$ within about 100 \Myr) than that of $\alpha$ and spans longer periods, during which more stars are formed. This is particularity visible in the C$_3$ group in the middle panel of \fig{features}. Therefore, the late part of the hook (\afe-decreasing) is naturally more populated than the early side (\afe-increasing). The over-densities of stars in the C groups characterize rapid evolutions with respect to nearby areas in the abundance plot, translating in a deficit of stars ``inside'' the hook of enrichment. We note that stars formed in starburst conditions like those of the features C are found on the high-\afe side in the \feh-\afe plane, but not necessarily in the age-\afe plane. This is due to the coeval formation of stars at higher \afe in smaller galaxies.

The overdensities of stars formed between C$_3$ and D$_1$ in the middle panel of \fig{features} constitutes the superposition of a number of distinct populations of various origins and thus do not trace any particular event.

\emph{Features D$_1$} and \emph{D$_2$} are characterized by a rapidly increasing \feh and a large spread in \afe, spanning almost the full range of the low-\afe sequence. The formation mechanism of these stars and its likelihood are presented in detail in \citetalias{Renaud2020b} and summarized here. Shortly before the last major merger, \vintergatan is fueled by two independent intergalactic gaseous filaments. The first connects the galaxy with the future last major merger. Because of the presence of these two massive galaxies and their outflowing material, the chemical content of this filament is slightly enriched. By fueling the galactic disk in the inner $\approx 5 \kpc$, it participates in the formation of stars at high metallicity ($\feh \gtrsim 0$). The second filament does not comprise another massive galaxy and thus is more metal-poor than the first. It is misaligned with the disk of \vintergatan and fuels a tilted outer structure as a flatten torus (inner radius $\approx 4 \kpc$, outer radius $\approx 10 \mh 15 \kpc$, depending on the azimuth) loosely resembling the geometry of a polar-ring. This tilted gaseous disk is however too diffuse to form stars. The early brushing passage of the last major merger tidally compresses the tilted structure and triggers the formation of the stars of the D$_1$ group in a non-bursty regime, i.e. without increasing \afe. This process occurs while star formation continues in the inner galaxy at higher metallicity ($\approx 9 \Gyr$, $z\approx 1.3$), which thus explains the detection of coeval stars with very different chemical compositions (see \fig{age} and the observations of \citealt{Feuillet2019}). At this time, the companion galaxy only plays the role of a trigger, but does not yet pollute neither the gaseous nor the stellar content of \vintergatan. Therefore, the D$_1$ stars have an in situ origin, but they inherit the peculiar kinematics of their natal decoupled tilted structure giving them halo-like kinematics (as the stars of comparable chemistry in the real Milky Way, \citealt{Haywood2018}). In particular, the D$_1$ group comprises a larger fraction of stars on retrograde orbits than the rest of the stars of the same age. With time, gravitational torques between the main disk and the tilted structure eventually align the two, but the structures remain chemically distinct for a few $100 \Myr$. The stars now being formed in the outer disk yield disk-like kinematics, but with a chemical enrichment in the continuity of that of the older component, making the D$_2$ feature. This star formation regime co-exists with that in the inner galaxy for $\sim 1 \Gyr$ until their respective enriched media mix (via radial flows, turbulence and fountains, see feature F). This ultimately populates the low-\afe sequence from both metallicity ends.

\emph{Feature E} has a similar enrichment history as the C groups (in term of slopes in \fig{features}), but at lower metallicity and 2 Gyr later. It is made of accreted stars formed in a relatively low mass galaxy, about 1 Gyr after the last major merger, although their chemical composition is close to that the most metal-rich stars from the last major merger. This confirms that the enrichment of lower-mass galaxies is slower than that of the massive ones, due to both a slower production of the metals (low SFR) and a less efficient retention (low escape velocity).

\emph{Feature F} is coeval with the transition from the high- to low-\afe sequences in the inner disk at high metallicity, but gathers low-metallicity stars found at the metal-poor end of the low-\afe sequence. These stars correspond to the secular evolution of the outer disk ($\gtrsim 5 \kpc$), while it aligns with the inner disk. Once the alignment done, young stars in the inner and outer disks have comparable kinematics but the two media are not efficiently mixed, keeping the stars at distinct locii in the abundance plot. The accretion of circum-galactic material onto the outer disk and the slow pollution of the inner disk from diffusion, turbulence and galactic fountains from the outer disk eventually homogenize the chemical compositions of the two components, at the end of feature F. This entire process is another indication that the low-\afe sequence assembles from two distinct and coeval channels. This is accompanied by a slight increase of \afe, but still capped in the low-\afe sequence. The physical origin of this increase is difficult to establish, but likely results from the peculiar enrichment regime of the tilting disk, where rapid star formation in self-enriching clumps occurs in the immediate vicinity of a low-density and low-metallicity coronal medium. This gas can condense onto the clumps, and the low densities can allow for the propagation of feedback to large scales. Both effects would effectively slow down the metallicity increase of the star forming regions. Such details on the enrichment of the tilting disk set the slope of the entire low-\afe sequence and how it connects to the inner disk. We note that the low-\afe sequence is significantly flatter (less decreasing with \feh) in \vintergatan than in observations (e.g. \citealt{Hayden2015}), possibly indicating a different fragmentation mode of the outer disk (leading to smoother structures in the real Galaxy), or a shorter depletion time (recall \sect{starbursts}). 

\emph{Features G$_1$, $G_2$} and \emph{$G_3$} are three low-mass galaxies accreted late by \vintergatan. These stars are still in the outer galaxy at high altitude ($|z| \approx 5\mh 10 \kpc$) and large radius ($R \approx 15 \mh 25 \kpc$) in our last snapshot. The high metallicities of these stars indicate that they formed recently in media largely polluted by outflows from massive galaxy(ies). Their small metallicity spread shows that, despite SFHs spanning about $1 \Gyr$ each, theses galaxies do not self-enrich significantly because of both a low SFR and their inability to retain outflows. These stars constitute the only low-\afe accreted material in our simulation but account for a very little fraction of the stellar component of \vintergatan. At a given age, outliers at the high end of the \afe distribution (e.g. features A and E) are made of accreted stars formed rapidly in a starburst regime in low-mass satellites. The converse statement is however not always true (e.g. features G), depending on whether the satellites assemble from almost pristine (A, E) or enriched (G) intergalactic medium.

\section{Discussion}
\label{sec:discussion}

\subsection{Observational detectability of the signatures of galaxy evolution}
\label{sec:detect}

\begin{figure*}
\centering
\includegraphics{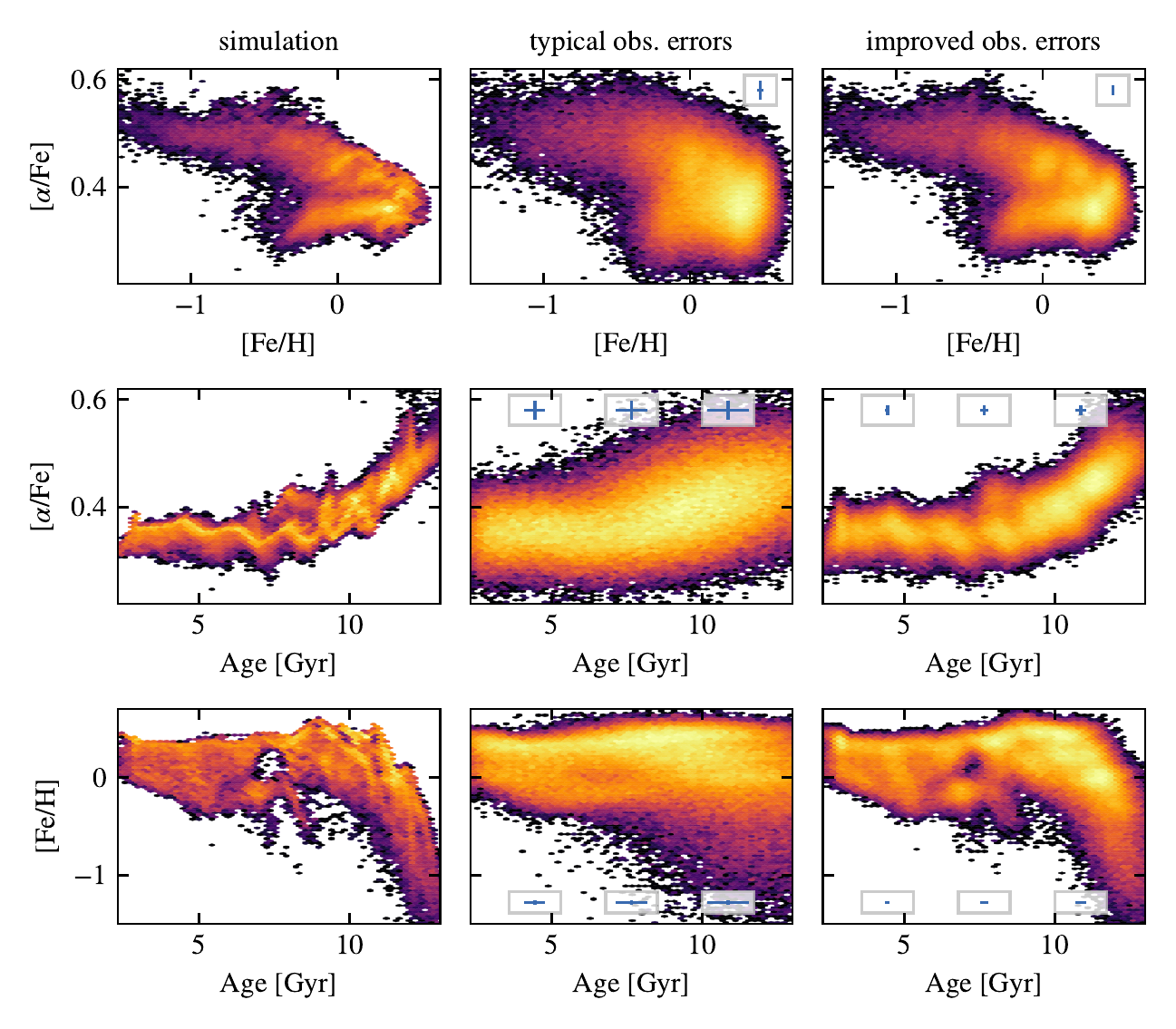}
\caption{Distributions of stars in the abundance plot, and as a function of age. The left column shows the output of the simulation, the central column includes typical observational errors on abundances and ages. Blue error bars illustrate the standard deviations of the Gaussian errors applied, shown at various ages in the middle and bottom rows. Most of the thin features are smoothed out when adding errors on the ages, but the general trends and even the metallicity bimodality remain visible, in particular through the deficit of $\sim 7 \Gyr$ old stars with solar metallicity. The right column includes reduced errors as estimates of the capacities of future instruments (see text for details). In that case, even the fine structures would be detectable.}
\label{fig:age_abundances_errors}
\end{figure*}

The simulation data presented here does not suffer from detection limits, observational uncertainties and is not restricted by any selection functions. For this reason, it is very unlikely to detect all the fine signatures discussed above in real data. To estimate at leading order the impact of these uncertainties, the central column of \fig{age_abundances_errors} shows the output of the simulation, stellar particle by stellar particle, convolved with Gaussian functions mimicking the typical amplitudes of observational errors. The standard deviations of the Gaussians are 0.03 dex for \afe, 0.05 dex for \feh (compatible with errors from APOGEE, \citealt{Nidever2014}), and an error of 0.18 dex for the ages, as typical of the isochrone matching technique \citep{Feuillet2016}. No luminosity limit nor selection function have been applied here, but we leave a similar analysis with more precise observational limitations and biases to a forthcoming paper.
 
With the first order convolutions presented here, we find, as expected, that the uncertainties on the ages have the most severe effects in blurring the signatures analyzed above. However, the global trends and the most contrasted features remain visible. In particular, we still detect the metallicity bimodality for $\sim 6\mh 7 \Gyr$ old stars corresponding to the simultaneous formation of stars in the low-\afe sequence at high and low metallicities, in the inner and outer disks respectively (features D and F in \fig{features}). Such a bimodality is a clear indication of a non-purely sequential assembly, but rather of several mechanisms active simultaneously. Therefore, if indeed detected in the real Galaxy, this feature could help distinguish between the formation scenario presented here and the two-infall model from \citet{Spitoni2019} and comparable alternatives \citep{Lian2020}. With sufficient confidence on the stellar ages, this could also help constrain the epoch of the last major merger. Considering other chemical elements, sensitive to different environmental conditions of star formation, could also provide additional insights to decode the formation history of the Galaxy \citep[e.g.][]{Jofre2019}.

The right column of \fig{age_abundances_errors} evaluates the detectability of finer structures when considering smaller uncertainties, for instance through the use of asteroseismology. This technique provides significant improvements on age determination, so far in specific regions of the Milky Way (e.g. with Kepler \citealt{Borucki2010}, and CoRoT \citealt{Baglin2006}). Here, we apply an error of 0.04 dex on the ages of all the stars (i.e. of the order of PLATO estimated values, \citealt{Rauer2014}), and also divide errors on abundances by a factor of 2, i.e. 0.015 dex and 0.025 dex for \afe and \feh respectively. Such improvements remain optimistic for the next observational programs like 4MOST \citep{deJong2019, Bensby2019}, but could become in reach of future generations of galaxy-wide data, for instance from large surveys from asteroseismology satellites. Such missions would thus unveil fine details, as those found in \vintergatan, which would greatly help testing the different galaxy formation scenarios.

\subsection{Effects of resolution on the hierarchical assembly}
\label{sec:resolution_mergertree}

By capturing the cold phase of the ISM (down to $\approx 10 \K$), our simulation resolves the cooling of gas in more numerous and lower-mass halos than at lower resolutions. This ultimately affects the hierarchical assembly and evolution of the most massive galaxies, although the long-term evolution of the mass of the main galaxy is roughly independent of resolution. For instance, by examining the stellar merger trees (i.e. excluding short-lived halos and unbound structures) at $z>4$, we find that \vintergatan counts about 3 times more galaxies of stellar mass $> 10^6 \Msun$ than the simulation of \citet{Renaud2017} using the same initial conditions but at $200 \pc$, $200\K$ resolution, and about 25 times more than the low-resolution ($500 \pc$, $10^4 \K$) case discussed in \citetalias{Agertz2020} (section 4.2).

At high resolution, the growth of massive galaxies is dominated by the repeated discrete accretion events in the form of many low-mass satellites, rather than the secular flow of intergalactic diffuse gas. Contrary to the dissipative gas, satellite galaxies can orbit the main one for several orbital times and induce significant tidal disruption. This affects the orbital motion of the massive galaxies and the transfer of angular momentum \citep{Kretschmer2020}, even changing the final orientation of the disk angular momentum vector by up to $90^{\circ}$ in cases mentioned above. In turn, this effect is important enough to delay the last major merger to $z \approx 1.2$ in this simulation, compared to $z\approx 2.0$ (i.e. a time delay of $\approx 1.8 \Gyr$) for the exact same initial conditions but a resolution of $200 \pc$. Yet, there is no evidence that convergence with resolution is reached at our resolution. However, by capturing smaller and smaller satellites with increasing resolution, the individual effects of these galaxies ought to become negligible on the final result (due to their low mass), which suggests that such a convergence exists.

\subsection{Capturing the width of the high-\afe sequence}
\label{sec:thickness}

Our results show that the merger activity is a direct driver of the chemical scatter of the old stellar population, in particular of the width in \afe of the high-\afe branch. In our simulation, this width is set by the superposition of a diversity of accreted galaxies, each providing their intrinsic scatter centered on different average chemical compositions, and by the signature of rapid enrichment during starburst events. We note that other simulations, with setups comparable to ours (i.e. cosmological zooms), can lead to significantly thinner \afe branches (e.g. \citealt{Brook2012, Stinson2013, Grand2018, Buck2020}). The same situation is also seen in our test run at low resolution (see figure 17 of \citetalias{Agertz2020}). Understanding these discrepancies would require in-depth explorations and tests beyond the scope of this paper. Nonetheless, we speculate below on three possible explanations.

First is the model-dependent variations in the injection and propagation of feedback, in particular the SN-Ia rates and yields. The importance of this effect is supported by the findings of \citet{Gibson2013} who report that the MUGS suite of simulations, which does not include a prescription for pre-supernova feedback (winds, radiative pressure etc. \citealt{Stinson2010}), yields chemical properties closer to observations than a more complete recipe with stronger feedback (MAGICC, \citealt{Stinson2013}), yet with important discrepancies on other global properties. The complex interplay of star formation, feedback, and the structure of the ISM makes these variations very complicated to understand and predict. We argue that our resolution of $\sim 20 \pc$ is sufficient to globally capture the propagation of enriched material in between the molecular clouds and in the intergalactic medium, but the very injection and early propagation of this material (from sub-parsec to parsec scales) remains out-of-reach of cosmological zoom simulations for now. Using the ad hoc setup of a single SN blast in a turbulent medium, \citet{Ohlin2019} demonstrated the difficulty of modeling the expansion velocity and extent of SN bubbles without capturing the details of the structure of the ISM at (at least) parsec scale. In addition, runaway stars having received a velocity kick from the parent star clusters can lead to the injection of feedback in remote locations and significantly alter the chemical enrichment, subsequent star formation and galactic outflows (\citealt{Andersson2020}). Quantifying the importance of these non-trivial effects on the galactic-scale chemistry requires a sub-parsec resolution to capture the porosity of the ISM, and a star-by-star description of the chemical polluters, which implies a collisional treatment of gravity to assess the dynamical evolution of binaries and multiple stellar systems \citep{Renaud2018b}. This would also improve the accuracy of the yields and rates of the type-Ia progenitors. The enormous numerical cost of such a setup, and the associated technical difficulties make this a very challenging task for the next generation of cosmological simulations.

Second, an explanation for different chemical scatters could lie in different merger histories and sensitivities of galaxies to reach the starburst regime. \fig{mm} shows that part of the chemical diversity of the simulated galaxy originates from the superposition of several galaxies, each with different histories and chemical enrichments. This is further amplified when considering minor mergers. Therefore, part of the width of the high-\afe sequence is directly linked to the merger history, the masses and number of satellites, and the epoch of their accretions. As noted in \sect{resolution_mergertree}, these details are sensitive to the numerical resolution of the simulation. In particular, the ability to capture the cooling of intergalactic gas has an important impact on the formation of low-mass galaxies, and thus on the accreted fraction. 

Third, simulations show that the starbursting nature of star formation in mergers derives from a turbulent compression of the ISM \citep{Renaud2014b}, which is fundamentally different than a mere increase of the star formation activity (i.e. an increase of the SFR in a non-starburst regime). Star formation in starbursts leads to different structures of the ISM, different turbulent excitations of the media, and different propagation of feedback than the normal star formation mode \citep{Renaud2019b}. However, capturing this change of regime requires simulations reaching high spatial resolutions and capturing the supersonic nature of the cold ISM ($\approx 10 \K$, $1\mh 10 \pc$ for a Milky Way-like galaxy). It is likely that \vintergatan still misses important details in the structure of the ISM and its susceptibility to merger-induced compression, and simulations at lower resolutions, in particular not resolving the thickness of the galactic disks with enough resolution elements, are very likely to miss the distinction between these two star formation regimes. In these cases, we suspect that mergers would still induce episodes of high SFRs (i.e. the production of more stars per unit time) due to an increased gas mass, but without capturing their starburst nature (i.e. the faster production of stars per unit gas, therefore short depletion times). As a consequence, such models would miss the hook-like enhancement of \afe (\sect{starbursts}), and thus an important widening agent of the \afe sequence.

\subsection{Possible effects of an active galactic nucleus}
\label{sec:agn}

Our simulation does not include a supermassive black hole and feedback from an active galactic nucleus (AGN). Although the center of the real Milky Way is not currently in an active phase, it is possible that it has been in the past, as suggested by the observation of the Fermi bubbles \citep{Guo2012}. In particular, gravitational torques induced by the passage of a companion galaxies are known to fuel galactic centers with gas \citep{Keel1985}, possibly increasing the accretion rate of central black hole, and turning the AGN on \citep{Ellison2011}. This is especially valid in the late stages of interactions (shortly before and at coalescence, \citealt{Hopkins2013b}). This regulates the post-merger star formation activity and is suspected to participate in temporary or definitive quenching \citep{Ellison2018}. Should such an event have happened, winds from the AGN would have enriched the intergalactic medium and stopped (or slowed down) star formation shortly after the last major merger. In the meantime, type-Ia SNe would continue to release iron in the ISM, while no new type-II SNe would explode in the absence of on-going star formation. Owing to this pollution being retained (or rapidly re-accreted) onto the disks, the \afe content of the ISM would then drop significantly more than in our model, and the metallicity would increase. After re-accretion, cooling and fragmentation of the ejecta of the AGN, this medium would then resume forming stars, likely in a non-starburst manner and thus remaining at low \afe and high \feh. Depending on how much the outer galaxy would be affected by AGN feedback, it is possible that the onset of star formation in the outer disk would not occur, or not at a metallicity as low as the one we found. In that case, the low-\afe sequence would yield a much narrower metallicity spread than in our model. However, the observation of $\approx 9 \Gyr$ old stars at both metallicity ends of the low-\afe sequence \citep{Feuillet2019} indicates that the redistribution of metallicities associated with AGN feedback and winds (if any) has not strongly affected the metal-poor content of the Galaxy at this epoch (see also \citealt{Ciuca2020}).

\section{Towards a comprehensive scenario for the formation of the Milky Way}
\label{sec:scenario}

\begin{figure*}
\centering
\includegraphics{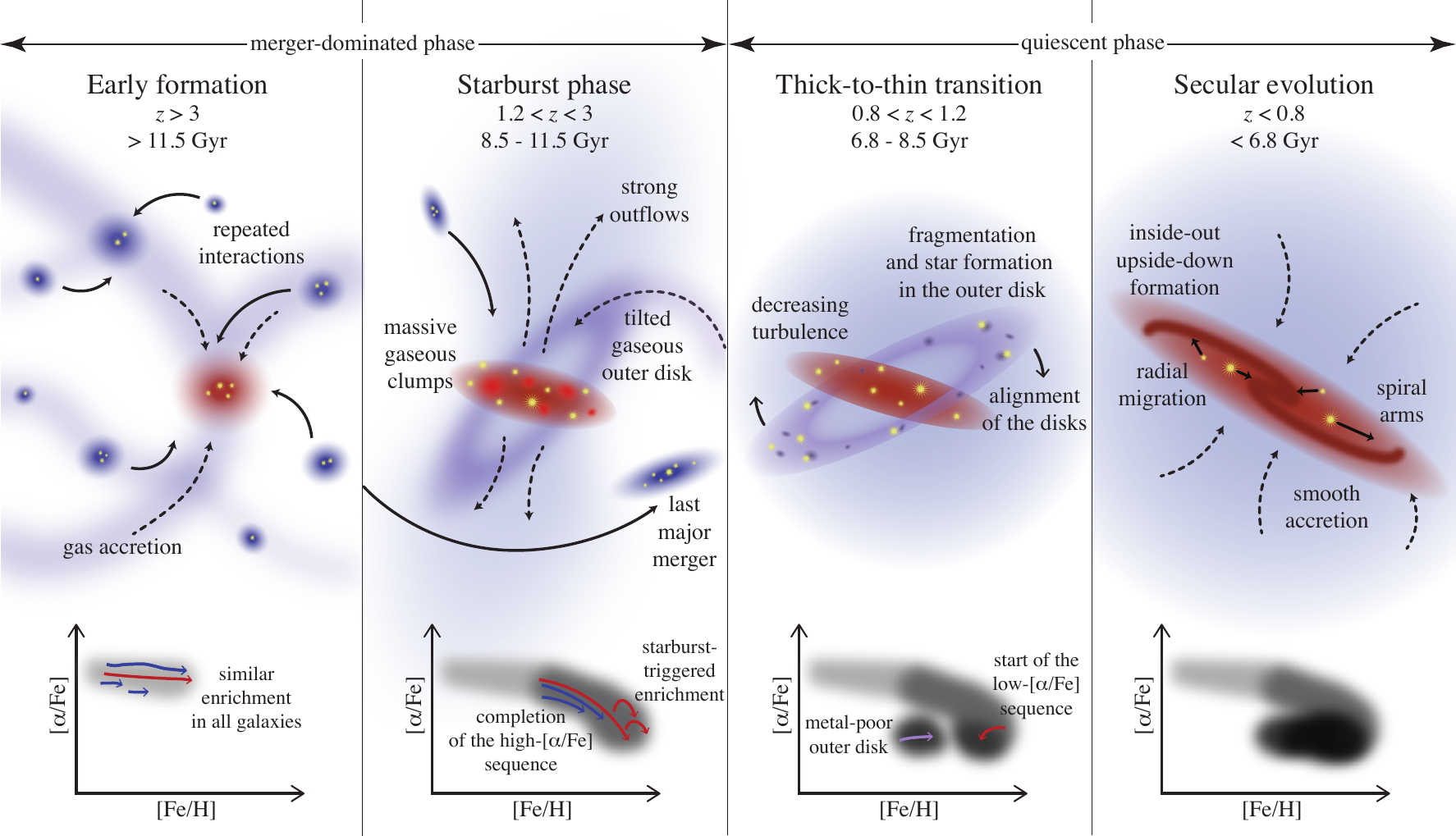}
\caption{Summary of the four main phases of our galaxy formation scenario. In the bottom panels, arrows indicate schematic evolutions of in situ population (red), the accreted stars (blue), and in situ stars formed in the initially tilted outer disk (purple, features D in \fig{features}, see \citetalias{Renaud2020b}).}
\label{fig:scenario}
\end{figure*}

We propose here a chronological description of the main phases and events along the evolution of \vintergatan, derived from the diagnostics presented here and in \citetalias{Agertz2020}. The main phases along galaxy evolution are illustrated in \fig{scenario}. We provide quantitative estimates of the key quantities, as derived from our simulation, but we remind the reader of the caveats inherent to our numerical approach, and the fact that the precise sequence of merger events likely differs between the simulation and the reality. Nevertheless, the compatibility of our results with observational data noted above and in \citetalias{Agertz2020} makes us confident that this qualitative scenario is applicable to the real Milky Way.

\subsection{Early formation and initial enrichment ($z > 3$)}

In the very first stages of galaxy formation ($z > 3$, $\gtrsim 11.5 \Gyr$), \vintergatan has a small stellar mass ($\sim 10^9 \Msun$) and grows relatively slowly. The accreted gas yields a low but rapidly rising \feh, and a mildly decreasing \afe. The intergalactic medium is accreted by a myriad of young galaxies, and the main galaxy itself does not particularly stand out yet. A major difference with the surrounding galaxies is its continuous increase of the in situ SFR ($> 10 \Msunyr$, \fig{sfr}). This induces the production of $\alpha$ elements by type-II SNe, later followed by Fe from type-Ia's \citep{Matteucci2012}. The still-shallow gravitational potential of the galaxy allows for an efficient release of this enriched material in the intergalactic medium (\fig{outflows}), which eventually contaminates all the nearby galaxies with the same chemical content. As a result, and because \vintergatan and all its progenitors at this early epoch have similar masses, the \afe sequence remains relatively narrow ($\approx 0.07 \dex$ for $\feh < -0.8$). In addition, the high levels of turbulence of the ISM imply an efficient chemical mixing and thus a small scatter in abundances within a given galaxy, at a given epoch (as noted observationally by \citealt{Ruchti2010}). Therefore, at low metallicity, the spread of the high-\afe branch mostly originates from the superposition of several distinct (yet similar) progenitor galaxies.

During this early phase, the disk is not yet in place (left panel of figure 1 of \citetalias{Agertz2020}) and the ISM is highly turbulent with disorganized dynamics. When major mergers occur (hence with low-mass galaxies of stellar masses about $10^{8\mh 9} \Msun$), the hydrodynamical response of the ISM to the perturbation is only mild \citep{Duc2013}. For instance, in the absence of a disk, tidal torques cannot efficiently fuel a nuclear starburst \citep{Keel1985, Barnes1991}. Hence, interactions and mergers only trigger modest and short-lived increases of the SFR, via shocks in the colliding interstellar media \citep{Jog1992} and tidal compression \citep{Renaud2014b}. It is instead the smooth, secular evolution of the galaxy via the accretion of intergalactic gas that is responsible for the overall rise of the star formation activity, thus mostly in situ (\fig{sfr}). Therefore, despite a globally increasing SFR, the star formation regime remains rather smooth, with long depletion times ($\sim 5\mh 10 \Gyr$, \fig{tdep}), such that the \afe content of the newly formed stars is mildly decreasing ($\sim 0.1 \dex \Gyr^{-1}$), while \feh rapidly increases ($\sim 1 \dex \Gyr^{-1}$).

At this epoch, the half-mass radius of the neutral gas of the galaxy is $\approx 2 \kpc$ (figure 6 of \citetalias{Agertz2020}), making the formation activity very concentrated. With time, gas cooling, transfer of angular momentum and large-scale tidal torques promote the formation of the thick disk \citep{Mo1998, Brooks2007, Kimm2011}.

\subsection{The starburst phase ($1.2 < z < 3$)}

At $z=2.7$ ($\approx 11.3 \Gyr$) and $z=2.2$ ($\approx 10.7 \Gyr$), \vintergatan experiences three major mergers (\fig{mm}), which rapidly multiplies its stellar mass by a factor $\approx 5$. About 45\% of this mass gain is the direct addition of the accreted material. However, now that a thick disk is in place, the interactions also efficiently trigger a starburst activity (up to $\approx 50\mh 60 \Msunyr$, \fig{sfr}), associated with short depletion times (\fig{tdep}) and accounting for the remaining 55\% of the increase of the stellar mass. These are the most intense star formation episodes the galaxy will ever experience, coinciding with the observed peak of cosmic SFR density \citep{Madau2014}. Such events participate in making \vintergatan the dominant galaxy of its volume. The repeated bombardment of the galaxy by (gas-rich) minor and major mergers maintains a highly turbulent ISM \citep{Renaud2014b} and a large vertical velocity dispersion in the stellar component \citep{DOnghia2016}. This favors the growth of the thick disk \citep{Robertson2006, Brook2007}, but also prevents thin structures from forming. This is in contrast with scenarios advocating for the formation of a thin disk first, dynamically heated later by mergers (see e.g. \citealt{Quinn1993} and a discussion in \citealt{Grand2020}). In our simulation, the disk cannot cool sufficiently to become globally thin before the end of the merger phase.

By bringing their own stellar populations to the main galaxy, the mergers necessarily widen the scatter in the chemical composition of the old stars. Furthermore, the starburst nature of the peaks of SFR induces the rapid enhancement of the \afe content of the ISM (\fig{tdep}), and thus creates the hook-shape structures in the abundance plot (like the features C in \fig{features}). A similar evolution takes place in the other massive galaxy of the group, the future last major merger (feature B in \fig{features}). The temporary boosts of the \afe content of the starbursting regions further widens the high-\afe sequence (to $\approx 0.2 \dex$ for $\feh = -0.4$). 

The type-Ia SNe formed in the early phase and during the starburst events progressively increase the \feh content of the intergalactic medium and the star forming ISM, until it reaches its maximum value ($z \approx 1.5$, $\approx 9.5 \Gyr$). In the abundance plot, this entire phase of galaxy evolution corresponds to the oblique section between the (approximately) flat part of the high-\afe branch at low metallicity, and its opposite extremity at high metallicity (\fig{mm}).

While the in situ formation reaches the most metal-rich end of the distribution, star formation in small satellite galaxies (to be accreted later) still proceeds at significantly lower metallicity (up to $\approx 1 \dex$ lower). In such low-mass systems, galaxy encounters trigger tidal disruption and quenching rather than starbursts, which keeps them on the lower part of the high-\afe sequence (\fig{selection}). The co-existence of these satellites and massive galaxies results in a very wide distribution in metals at a given age, during this period (\fig{isfrac_contours}). Therefore, in the high-\afe sequence, the in situ material is preferentially found at the high-\afe end and at high metallicity (for a given age), while the stars at a lower \afe are mostly accreted from low mass satellites and thus significantly more metal-poor. 

During this phase, the gas fraction of \vintergatan's disk is about $30\mh 40\%$, which favors gas-driven large scale disk instabilities \citep{Agertz2015b, Romeo2020} and the formation of massive clumps, similar to those observed in $z\approx 1-3$ disks \citep{Wuyts2012}\footnote{With a lower gas fraction, the stellar component would dominate the dynamics of the disk and the instabilities to develop would resemble spiral arms and giant molecular clouds along them, rather than massive clumps.}. Such an irregular structure of the ISM with sharp density contrasts is reinforced by the repeated tidal disturbance of the disk and enhanced compressive turbulence by interactions and mergers \citep{Renaud2019b}. All together, this increases the porosity of the ISM, allowing stellar feedback to couple to large scales ($> 1 \kpc$, \citealt{Ohlin2019, Martizzi2020}). Thus, galactic winds efficiently launch enriched material from the clumpy, turbulent disk \citep{Hayward2017}. The ejecta mix with low metallicity gas in the circumgalactic and the intergalactic media. Interactions between the outflowing material and the galactic corona medium promote the condensation of low-metallicity gas onto the disk (Figure 5 of \citetalias{Agertz2020}, see also \citealt{Fraternali2013}) which eventually becomes star forming, hence at a lower metallicity than the ISM at the epoch of ejection. 

Without such sharp density contrasts around the star forming regions, the enriched gas would be more efficiently trapped within the galaxy, i.e. it would not mix with more metal-poor gas, and would eventually form more metal-rich stars. An efficient propagation of feedback to intergalactic scales and its mixing with metal-poor content is thus key in capping the self-enrichment of the galaxy. In addition, low-metallicity medium from the outer disk eventually reaches the inner galaxy and effectively lowers the metallicity of the stars formed after the last major merger. Combined with a production of metals slowing down due to a decreasing SFR after the starburst activity ceases, this explains that the galaxy reaches its maximum metallicity at $z\approx 1.5$. This is about $4\mh 5 \Gyr$ earlier than observed by \citet{Feuillet2019} in the solar neighborhood ($7\mh 9 \kpc$). However, the radial dependence of the metallicity suggests that, like in \vintergatan, the maximum metallicity could be reached significantly earlier in the inner galaxy (as seen in \citealt{Ciuca2020}, see their figure 9 which is in remarkable agreement with our conclusions).

During the first passages of galaxies, the disk usually gets tidally truncated, making it more compact \citep{Gnedin2003}. However, the coalescence of the galaxies and the re-accretion of tidal debris increase the total angular momentum of the disk (most notably in the case of prograde encounters, \citealt{Duc2013}) making the disk more extended on the long term \citep{Peschken2017}. As a result, the half-mass radius of the atomic gas slightly increases, but in a complex non-monotonic fashion, during the merger-dominated phase (figure 6 of \citetalias{Agertz2020}).

\subsection{The end of the major merger phase and the transition from thick to thin disk ($0.8 < z < 1.2$)}

The galaxy involved in the last major merger (first passage at $\approx 9.5\Gyr$, $z=1.3$ and final coalescence at $\approx 8.7\Gyr$, $z=1.2$) causes similar effects to the previous mergers. By definition, it marks the end of the merger-dominated growth phase of \vintergatan, which allows for a less violent subsequent evolution. 

Via tidal compression, the passage of the last major merger triggers the formation of star forming clumps in the outer tilted disk (features D in \sect{features} and \citetalias{Renaud2020b}). This new stellar population is found at the low-metallicity end of the low-\afe sequence. The metallicity of this structure increases and reaches its maximum within $\approx 1.2 \Gyr$, likely for the same reason discussed in the previous section, i.e. an efficient ejection of polluted material from a clumpy medium (figure 4 of \citetalias{Renaud2020b}). In \citetalias{Renaud2020b}, we show that this precise chain of events is likely to happen in Milky Way-like galaxies. Should star formation in the outer disk have proceeded with shorter depletion times, i.e. in a starburst mode induced by a more penetrating passage of the last merger for instance, the \afe composition of these stars would have been higher (\sect{starbursts}), placing them closer to the high-\afe sequence at the same \feh which comprises stars about $3 \Gyr$ older. A similar effect is expected from the formation in rapidly self-enriching massive dense clumps ($\sim 10^9 \Msun$), as found galactic-wide in the isolated galaxy model of \citet{Clarke2019}. This would then have changed the slope of the low-\afe sequence in the abundance plot, possibly making it more similar to the observed shape (e.g. \citealt{Hayden2015}). In \vintergatan, star formation in the outer disk proceeds in relatively small gas clumps ($\sim 10^{6\mh 7} \Msun$) formed by the brushing passage of the last major merger, which results in the non-enhancement of the \afe content of these stars, and an \afe composition approximately independent of the metallicity in the low-\afe branch. The slope of the low-\afe sequence is thus likely strongly dependent on the details of the physics of star formation in the outer disk (see also \citealt{Ciuca2020}).

While formation in the outer disk starts as early as $9 \Gyr$ ago (compatible with the estimates of \citealt{Ciuca2020}), we note that the majority of stars in the low-metallicity part of the low-\afe sequence form $2\mh 4\Gyr$ later, i.e. with ages of $5\mh 7 \Gyr$ (feature F and later in \fig{features}). This is slightly earlier, but yet in qualitative agreement, with the results of \citet{Wu2019} and \citet{Lian2020} who found that the bulk of star formation with this chemical composition occurred about $5 \Gyr$ ago.

With the onset of star formation in the tilting disk, the in situ formation of stars proceeds along two distinct channels, which translates into a clear bimodality in metallicity. Rather than the sudden decrease of the metallicity across the entire galaxy (e.g. due to a gas accretion in the two infall scenario, \citealt{Spitoni2019}), the bimodality in \afe thus results (at least initially) from the onset of a new formation channel in the outskirts of the galaxy. It eventually connects to the more metal-rich population in the inner disk which makes the bulk of the low-\afe sequence. Therefore, the key to the existence of the \afe bimodality and its properties (chemistry, epoch) are directly linked to the transition from a merger-dominated growth phase to a secular evolution of the main galaxy.

After the final coalescence of the last major merger, gravitational torques between the inner and the outer disks align them together, marking a rapid radial growth of both the gaseous and stellar disks (figures 6 and 7 of \citetalias{Agertz2020}). The scale radius of the stellar disk changes from $\approx 2 \kpc$ for the old high-\afe component (formed during the merger-dominated phase) to $\approx 4.4 \kpc$ for the young, post-merger, low-\afe population (figure 9 of \citetalias{Agertz2020}), in agreement with observations of a thin disk being more extended than the thick disk \citep{Bensby2011}. Stars formed early in the tilted outer disk retain a signature of their peculiar origin through dynamically hot properties, like a high vertical velocity dispersion and a large fraction of retrograde orbits (see figures 6 and 7 of \citetalias{Renaud2020b}), and share kinematic properties of halo stars. This is compatible with the observational results on chemistry and kinematics from \citet{Ciuca2020} and \citet{Bonaca2020}. However, the interpretation of quenching induced by the last major merger proposed by \citet{Bonaca2020} does not match our simulation. Once the alignment of the inner and outer disks is done ($\approx 5 \Gyr$, $z = 0.5$), the ISM of the two structures mixes more efficiently, which effectively lowers the metallicity of the inner part of the disk ($\lesssim 5\kpc$).

\subsection{Late secular evolution ($z < 0.8$)}

At this point of galaxy evolution, the accreted material constitutes 40\% of the total stellar mass, but this fraction drops with time due to the rarefaction of merger events and the decrease of the average mass of the galaxies being accreted. (The final fraction of accreted stellar mass is $28\%$, of which $35\%$ originate from the last major merger alone, i.e. $10\%$ of \vintergatan.) Late minor mergers only have a limited impact on \vintergatan itself. They can induce mild tidal deformations such as disk warps \citep{Ibata1998, Laporte2018, Skowron2019} and rings \citep{Newberg2002, Morganson2016}, as observed in the real Galaxy. The tidal disruption of incoming satellites leaves stellar and gaseous streams that slowly disperse under the effect of galaxy substructures (e.g. \citealt{Carlberg2011, Sandford2017}). The stars accreted late from dwarf satellites are thus preferentially expected to have halo-like kinematics and distinct chemical composition from the bulk of the Milky Way. If formed late, i.e. in the IGM polluted by the main galaxy, these satellites would rather share the chemical composition of the outflows. On the contrary, if formed early (as thought in the case of the satellites in the Local Group, \citealt{Tolstoy2009}), then an extended star formation history would confer them a low \afe \citep{Unavane1996}, while an early formation episode rapidly quenched (e.g. by re-ionization) would yield a high \afe.

The end of repeated major interactions appears as a necessary condition for the formation of the thin disk, but it is not a sufficient one, since the turbulence support of the ISM also needs to decrease (which is likely linked to the decrease of the gas fraction, \citealt{Fisher2014}). Therefore, the formation of the thin disk does not necessarily coincide with that of the low-\afe sequence, but the onsets of the two events share some common necessary physical conditions. The decrease of both the tidal excitation and the level of turbulence effectively shrinks the scale-height of the gas disk, leading to star formation being confined in a thin disk (although the disks are naturally subject to flaring at large radii, see figure 12 of \citetalias{Agertz2020}, \citealt{Struck2019}). This evolution also translates into a decreasing vertical velocity dispersion for the young populations (figure 14 of \citetalias{Agertz2020}). With time, scattering between stars gives the stellar disk its exponential profile (figure 9 of \citetalias{Agertz2020}, \citealt{Elmegreen2013}). 

In the central region ($\lesssim 2 \kpc$, figure 9 of \citetalias{Agertz2020}), the bulge gathers stars of all ages and all chemical compositions, with a non negligible contribution of the accreted population (\fig{isac_position}). This is a relic of early formation in a non-axisymmetric medium (before the formation of the disks, $z \gtrsim 4$), of the overall inside-out formation of the disks \citep{Chiappini2001, Pilkington2012, Frankel2019}, and of the complex superpositions of several populations in the galactic center (\citealt{Zoccali2017, Thorsbro2020}, including accreted material with misaligned angular momentum).

After the last major merger, the gas fraction in the disk has decreased to about $20\%$. The stellar component now dominates disk dynamics, and large-scale instabilities are no longer set by the ISM. The formation of massive clumps ceases and is replaced by spiral-like instabilities (figures 1, 5 and 8 of \citetalias{Agertz2020}). The radially-extended thin disk resulting from the alignment of the inner and outer structures promotes the development of long spiral arms. The spirals strengthen with time, and dynamical interactions with stars induce radial migration (figure 11 of \citetalias{Agertz2020}, \citealt{Sellwood2002, Minchev2013}). 

A strong bar does not form in our simulation, possibly because of energy dissipation due to a slightly too high gas fraction, an important stabilization by the bulge or the halo \citep{Debattista2000, Kataria2018}, or a too strong gas accretion \citep{Kraljic2012}. This absence causes differences between \vintergatan and the real Galaxy. For instance, we would expect a bar to trigger star formation at its extremities \citep{Renaud2015d, Motte2018}, to reduce star formation inside the bar \citep{Longmore2013b, Emsellem2015}, and to fuel nuclear star formation in the very center where the gas accumulates. A bar would also affect the overall kinematics of the disk, in particular by introducing resonances, altering radial migration and fueling gas toward the nucleus \citep{Lynden1972}. Exploring these points requires a series of dedicated simulations, which is beyond the scope of this paper.

\section{Summary and conclusion}

Using the \vintergatan cosmological zoom simulation, we have analyzed the role and contributions of galaxy interactions and mergers on the assembly of a Milky Way-like galaxy. This has led us to establish a comprehensive formation scenario for the Galaxy, compatible with observational data, as presented in \sect{scenario} and summarized in \fig{scenario}. Our main results are:
\begin{itemize}
\item Accreted stars contribute to about half of the high-\afe sequence. The coeval stars formed in situ are preferentially found on the high-\afe side of this sequence.
\item The different star formation histories and abilities to retain enriched material of the progenitors of the galaxy lead to very little overlap of the accreted and in situ populations in the \feh-\afe plane. Furthermore, accreted stars are almost totally absent of the low-\afe sequence. This distinction is blurred by observational uncertainties and is likely not discernible with modern techniques, but will be with near future improvements.
\item Once a galactic disk is in place, its response to major mergers leads to repeated starbursts. The \afe content of the stars formed during such events of short depletion times is enhanced with respect to the more quiescent star formation regime.
\item The passage of the last major merger galaxy near a diffuse gaseous disks in the halo triggers star formation at the metal-poor end of the low-\afe sequence (see \citetalias{Renaud2020b} for details). This is concomitant with the transition from high- to low-\afe at high metallicity in the inner galaxy. With time, this outer disk aligns with the inner one, but some of its stars retains halo-like kinematics.
\item The transition from thick to thin disk formation is allowed by the end of the merger-dominated growth phase, i.e. the end of the starburst activity, tidal heating, and merger-driven turbulence. This geometrical transition is accompanied with a transition to colder kinematics and the onset of the low-\afe sequence.
\item At the same epoch, the decrease of the gas fraction in the disk replaces large-scale instabilities in the form of massive gas clumps with smoother and finer, spiral-like structures. This modification of the porosity of the ISM and of the density contrasts around star forming regions leads to the coupling of feedback to larger scales. Consequently, the metallicity reaches its maximum at this epoch. Subsequent condensation of the coronal medium ultimately induces a mild yet significant decrease of the metallicity of the newly formed stars.
\item Therefore, the cessation of the merger activity is the key event allowing for the onset of the properties associated with the thin disk, namely no longer high \afe from starbursts, low and decreasing SFR, reduced gas turbulence, low gas fraction, low vertical velocity dispersion, formation of spirals, radial migration etc. This effectively marks the transition between the two main phases of galaxy formation: a violent growth dominated by mergers ($z \gtrsim 1$), followed by a smoother secular evolution.
\end{itemize}

These results suggest that the observed chemical bimodalities directly results from the merger history of the Milky Way, in particular from the properties of its last major merger. In that respect, we predict that disk galaxies with a mass high enough to be able to self-enrich, gravitationally dominate their surroundings and clear them from satellites at $z\approx 1\mh 2$ would show similar chemical patterns as the Milky Way, including the bimodality in \afe. Conversely, galaxies which experience major mergers at different epochs, and with different mass ratios would likely not exhibit the same chemical properties, and may even lack bimodal chemical distributions.

In our scenario, the transition from thick to thin disk is equivalent to that from high to low \afe, kinematically hot to cold, radially compact to extended, and occurs in a smooth manner, as suggested by the observations (see e.g. \citealt{Bensby2014, Feuillet2019} on the ages, \citealt{Reddy2003, Nordstrom2004, Hayden2020} on the turbulent, kinematically hot and old thick disk, \citealt{Bensby2011} on the radially compact thick disk, and \citealt{Bovy2016} on the smooth transition between the two groups). Due to the stochasticity of the merger history in cosmological simulations, it is possible to imagine a very similar formation scenario, but with different timing for the last major merger. In particular, with sufficient time between the penultimate and the last major merger, and with a low enough gas-fraction, it could be possible for the Milky Way disk to cool and initiate star formation in the thin disk (at low \afe) before the last major merger. The last interaction would then pause this new mode by temporarily re-exciting orbits and turbulence. As a consequence, around the epoch of the last major merger, the thin and thick disk (and the low- and high-\afe sequences) would yield stars of similar ages, as suggested by the overlap of the age distributions noted in the solar neighborhood \citep{Silva2018}. We will explore this hypothesis with forthcoming dedicated simulations.

\section*{Acknowledgements}
We thank the referee for their constructive report, as well as Leandro Beraldo e Silva, Avishai Dekel, Carme Gallart, Paula Jofr\'e, Sergey Khoperskov, Jianhui Lian, Ted Mackereth, and Paola Di Matteo for their interesting and enthusiastic input. FR, OA, EA and MR acknowledge support from the Knut and Alice Wallenberg Foundation. OA acknowledges support from the Swedish Research Council (grant 2014-5791). TB was funded by the grant 2018-04857 from the Swedish Research Council. DF was supported by the grant 2016-03412 from the Swedish Research Council. 

\section*{Data availability}
The data underlying this article will be shared on reasonable request to the corresponding author.

\bibliographystyle{mnras}
\bibliography{biblio}

\begin{appendix}
\section{Spatial distribution of the accreted and in situ populations}
\label{sec:position}

\begin{figure}
\centering
\includegraphics{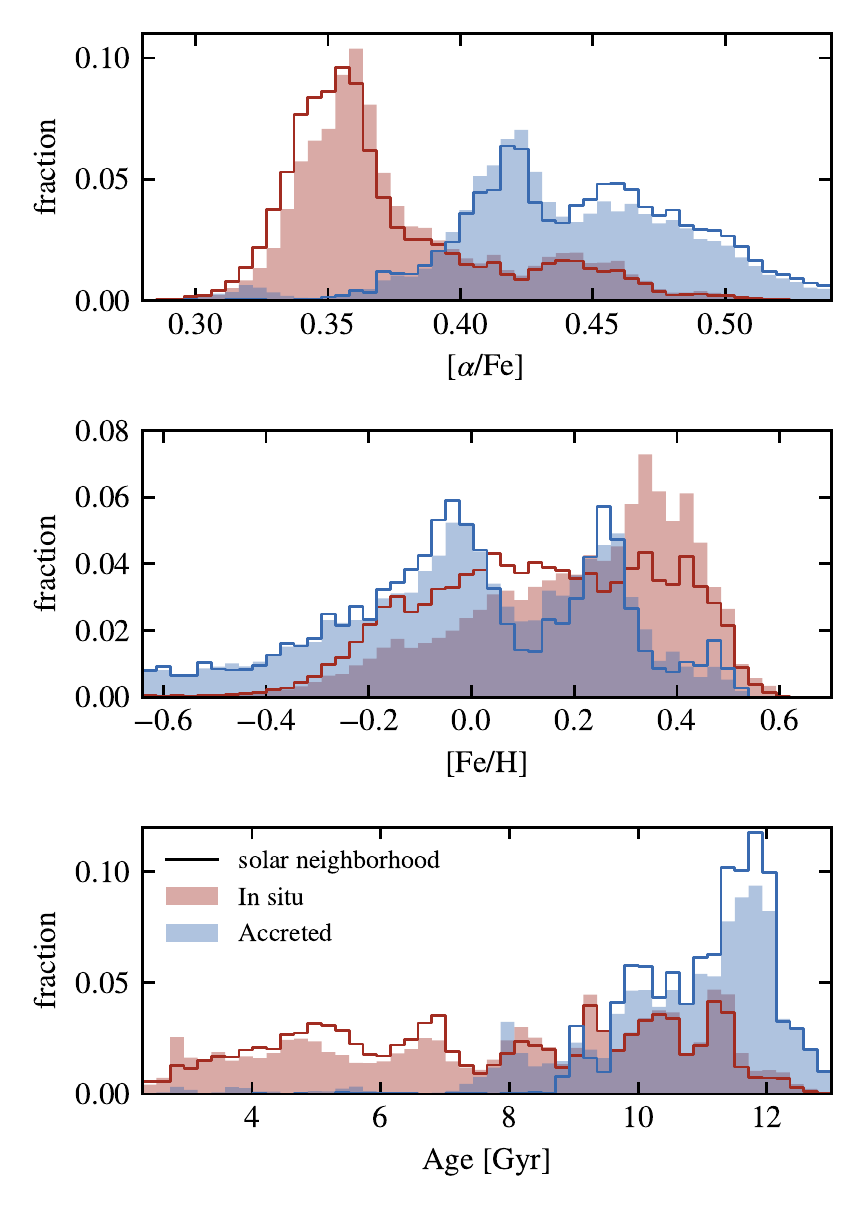}
\caption{Normalized distributions of the in situ and accreted populations for the entire galaxy (filled histograms, as in \fig{isac_histo}) and for the solar neighborhood only (solid lines, $7 < R < 9 \kpc$).}
\label{fig:solar_histo}
\end{figure}

\fig{solar_histo} repeats the analysis of \fig{isac_histo}, but focusing on the region of the modeled galaxy equivalent to the solar neighborhood ($7 \kpc < R < 9 \kpc$). We considerer here the stars found in the solar neighborhood on the final snapshot of the simulation, i.e. including those which formed elsewhere and having migrated.

The distribution of \afe of stars in the solar neighborhood is comparable to the rest of the galaxy, with only a slight excess at the low-\afe end. In term of \feh, stars formed in situ and found now around the Sun have a flatter distribution than the overall galaxy, i.e. more metal-poor stars ($\feh \sim -0.2 \mh 0.2$), and less metal-rich stars ($\feh \sim 0.2 \mh 0.5$), due to the negative radial gradient in \feh (\citetalias{Agertz2020}). This effect is not visible for the accreted stars, since potential metallicity gradients in the progenitor galaxies are likely to be erased during the merger. This confirms the efficient mixing of the accreted population within the galaxy noted above. The similarities between the solar neighborhood populations and the entire galaxy for the accreted stars is compatible with the observational results of \citet{Nidever2014} and \citet{Hayden2015} who found the metallicity distributions to be independent of location in the high-\afe sequence (i.e. traced by our accreted population, recall \fig{isac_histo}).

\begin{figure*}
\centering
\includegraphics[width=0.9\textwidth]{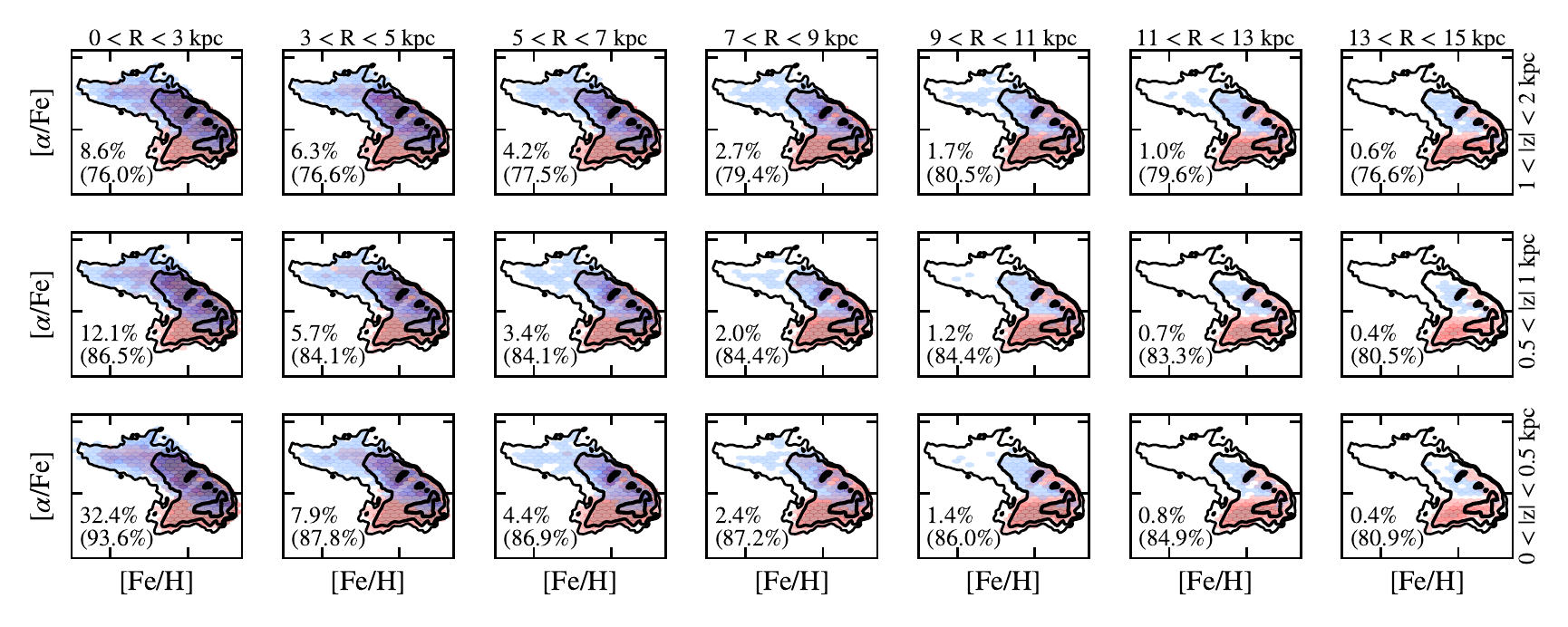}
\includegraphics[width=0.9\textwidth]{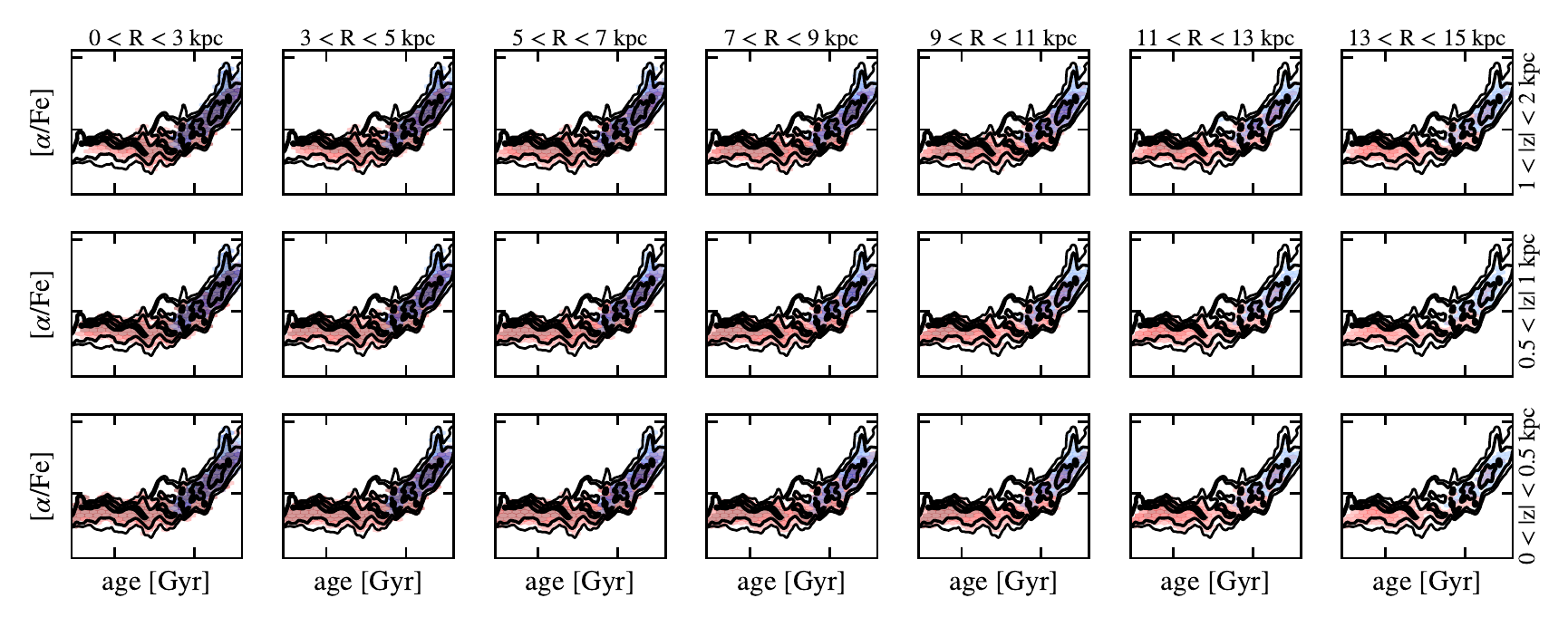}
\includegraphics[width=0.9\textwidth]{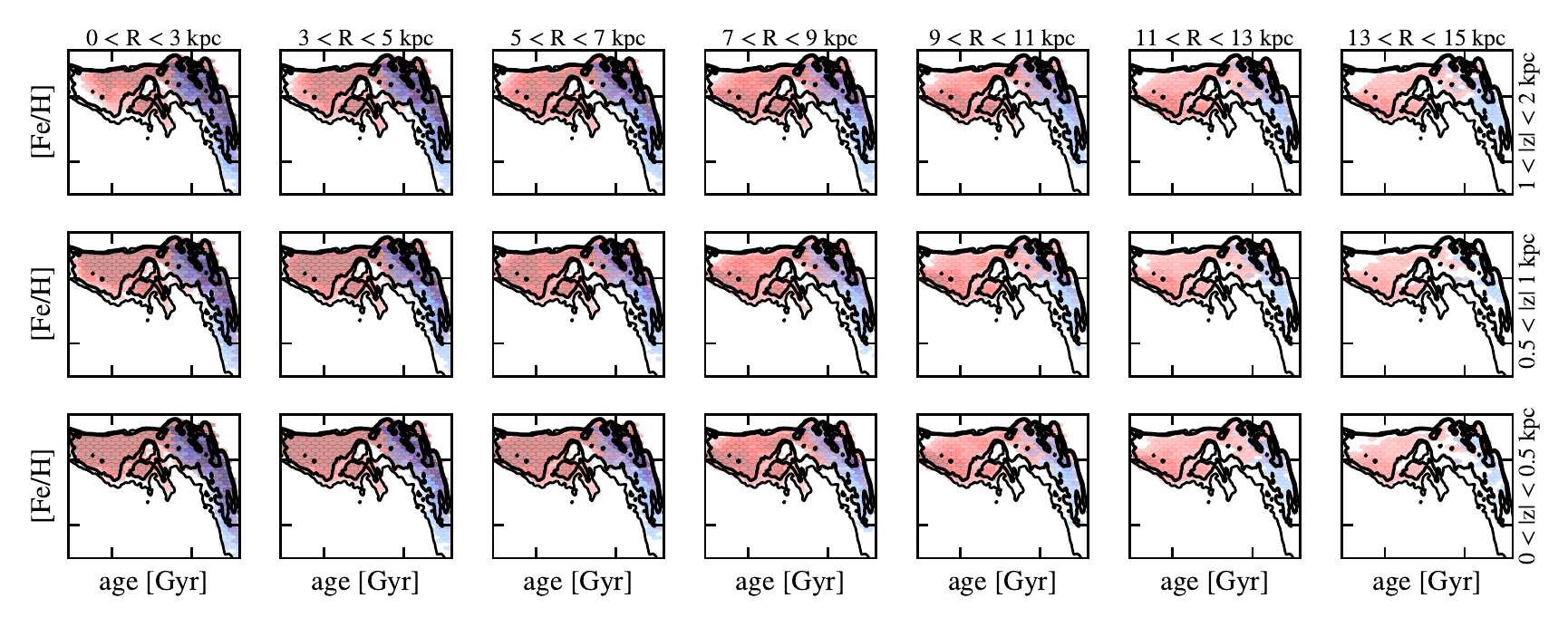}
\caption{Distributions of the in situ (red) and accreted stars (blue) in radial and vertical position bins. All the distributions are normalized independently of each others. Contrarily to \fig{isfrac_contours}, this does not show the fraction of in situ stars, but rather the distributions of accreted and in situ populations superposed to each other. The black lines indicate the contours encompassing 0.1, 1 and 10\% of the entire stars. The two percentages in the bottom-left corner of each panel indicate the fraction of stars in the sub-volume with respect to the entire volume covered ($R < 15 \kpc$ and $|z|< 2 \kpc$), and the fraction of in situ stars in the panel (in parentheses).}
\label{fig:isac_position}
\end{figure*}

\fig{isac_position} shows the distributions of in situ and accreted populations in different radial and vertical bins. The rare areas of overlapping populations in this plane (in the high-\afe branch, recall \fig{isfrac_contours}) are exclusively found in the inner disk (up to $7\mh 9 \kpc$) and become rarer with decreasing height. This corresponds to the overlap of the bulge (i.e. overlap of old stars of various origins) and the innermost parts of the disks which host on-going in situ star formation, as well as the old in situ component formed in the thick disk (\citetalias{Agertz2020}). Apart from the innermost region ($< 3 \kpc$) where nuclear star formation produces significant amounts of young, metal-rich, low-\afe in situ stars (at the end of the simulation, the inner $3 \kpc$ host 60\% of the total SFR of the galaxy), the fraction of in situ stars varies at most by a few percents with the height (up to $|z| = 2 \kpc$). 

Adding the information on the age, we find that the outer disk ($\gtrsim 11 \kpc$) has a deficit of intermediate-age stars ($\approx 7\mh 10 \Gyr$), marking a clear dichotomy between the old, accreted population ($\gtrsim 10 \Gyr$) with some contribution of the high-\afe in situ stars, and the young, low-\afe stars ($\lesssim 7 \Gyr$) almost exclusively of in situ origin. All the other regions show comparable distributions in \afe to that of the entire galaxy, indicating a efficient blend of $\alpha$ elements and iron, for both the in situ and accreted stars.

Most of the old stars lie in the central regions, irrespective of their origin. This is also where the most metal-poor ($\feh \lesssim -1$) and the most metal-rich ($\feh \gtrsim 0$) stars are found, while the outer disk only gathers stars of intermediate metallicities ($\feh \approx -0.2$).

\end{appendix}

\end{document}